\documentclass[aoas]{imsart}

\RequirePackage{amsthm,amsmath,amsfonts,amssymb}
\RequirePackage[authoryear]{natbib}
\RequirePackage{graphicx}%% uncomment this for including figures

\RequirePackage{booktabs,array}

\usepackage{xcolor}

\startlocaldefs
\theoremstyle{plain}

\newtheorem{theorem}{Theorem}[section]
\newtheorem{lemma}[theorem]{Lemma}
\theoremstyle{definition}
\newtheorem{property}{Property}
\endlocaldefs

\def\bfg{{\ensuremath{\bf g}}}

\def\bfx{{\ensuremath{\bf x}}}

\def\bfX{{\ensuremath{\bf X}}}

\def\bfzero{{\ensuremath{\bf 0}}}

\def\bftheta{{\ensuremath\boldsymbol{\theta}}}

\def\bfeta{{\ensuremath\boldsymbol{\eta}}}
\def\bfalpha{{\ensuremath\boldsymbol{\alpha}}}
\def\bfbeta{{\ensuremath\boldsymbol{\beta}}}
\def\bfgamma{{\ensuremath\boldsymbol{\gamma}}}

\def\bfrho{{\ensuremath{{\boldsymbol{\rho}}}}}

\def\bfSigma{{\ensuremath\boldsymbol{\Sigma}}}

\begin{document}

\begin{frontmatter}
%%%%%%%%%%%%%%%%%%%%%%%%%%%%%%%%%%%%%%%%%%%%%%
%%                                          %%
%% Enter the title of your article here     %%
%%                                          %%
%%%%%%%%%%%%%%%%%%%%%%%%%%%%%%%%%%%%%%%%%%%%%%
\title{A Novel Tool for Evaluating Effect Modification in Older Adults with ADRD Using Medicare Claims}
%\title{A sample article title with some additional note\thanksref{T1}}
\runtitle{A Novel Analytical Tool for Evaluating Effect Modification}
%\thankstext{T1}{A sample of additional note to the title.}

\begin{aug}
%%%%%%%%%%%%%%%%%%%%%%%%%%%%%%%%%%%%%%%%%%%%%%%
%% Only one address is permitted per author. %%
%% Only division, organization and e-mail is %%
%% included in the address.                  %%
%% Additional information such as            %%
%% identifying the corresponding author must %%
%% be included in in the Acknowledgments     %%
%% section if necessary.                     %%
%% ORCID can be inserted by command:         %%
%% \orcid{0000-0000-0000-0000}               %%
%%%%%%%%%%%%%%%%%%%%%%%%%%%%%%%%%%%%%%%%%%%%%%%
\author[A]{\fnms{Yilin}~\snm{Zhang}\ead[label=e1]{yilinzhang@som.umaryland.edu}},
\author[B]{\fnms{Michelle}~\snm{Shardell}\ead[label=e2]{mshardell@som.umaryland.edu}},
\author[C]{\fnms{Jason}~\snm{Falvey}\ead[label=e3]{jfalvey@som.umaryland.edu}},
\author[A]{\fnms{Rozalina}~\snm{McCoy}\ead[label=e4]{Rozalina.McCoy@som.umaryland.edu}},
\author[E]{\fnms{Elizabeth}~\snm{Stuart}\ead[label=e5]{estuart@jhsph.edu}}
\and
\author[A]{\fnms{Chixiang}~\snm{Chen}\thanks{\textbf{Corresponding author}}\ead[label=e6]{chixiang.chen@som.umaryland.edu}}
%%%%%%%%%%%%%%%%%%%%%%%%%%%%%%%%%%%%%%%%%%%%%%
%% Addresses                                %%
%%%%%%%%%%%%%%%%%%%%%%%%%%%%%%%%%%%%%%%%%%%%%%
\address[A]{University of Maryland Institute for Health Computing\printead[presep={,\ }]{e1,e4}}

\address[B]{Institute for Genome Sciences Health Sciences Facilities III,University of Maryland\printead[presep={,\ }]{e2}}

\address[C]{Department of Physical Therapy and Rehabilitation Science,University of Maryland Baltimore\printead[presep={,\ }]{e3}}

\address[D]{Department of Epidemiology and Public Health,University of Maryland Baltimore\printead[presep={,\ }]{e6}}

\address[E]{Department of Biostatistics, Johns Hopkins Universit\printead[presep={,\ }]{e5}}
\end{aug}

\begin{abstract}
Studying consequences following baseline exposures has become increasingly important for advancing comparative effectiveness research using real-world data. This case study evaluates the impact of hospital-acquired conditions (HAC) during hospitalization for hip fracture on post-discharge recovery trajectories among older adults living with Alzheimer’s Disease and Related Dementia, a population particularly vulnerable to high post-hospital mortality. To appropriately account for truncation of recovery trajectory due to death and to explore heterogeneity in effect modification by patient demographics, we introduce a novel pseudo data-based robust (PD-Robust) analysis strategy, accompanied by an R package and detailed usage guidance to inform real data analysis. Grounded in an interpretable estimand via principal stratification under principal ignorability and a structural working model, PD-Robust accommodates truncation by death, provides {model diagnosis and robustness check against assumption violation}, and facilitates the characterization of patient profiles among the principal stratum. Applied to Medicare claims data, where better recovery is defined as more days at home (DAH) over six months post-discharge, PD-Robust reveals heterogeneity in HAC effects, with males under the age of 85 years as a high-risk subgroup experiencing up to $23$ fewer DAH, comparing HAC to no HAC. This exceeds the $8$-day threshold regarded as clinically meaningful difference in DAH due to any exposure. Moreover, simulation studies further demonstrate that PD-Robust achieves low estimation bias and accurate statistical inference, supporting its utility in real-world data applications. 
\end{abstract}


\end{frontmatter}
%%%%%%%%%%%%%%%%%%%%%%%%%%%%%%%%%%%%%%%%%%%%%%
%% Please use \tableofcontents for articles %%
%% with 50 pages and more                   %%
%%%%%%%%%%%%%%%%%%%%%%%%%%%%%%%%%%%%%%%%%%%%%%
%\tableofcontents

%%%%%%%%%%%%%%%%%%%%%%%%%%%%%%%%%%%%%%%%%%%%%%
%%%% Main text entry area:

\section{Introduction}
\label{sec:intro}

\subsection{ Motivation from real-world data}

Evaluating clinical outcomes after the initiation of interventions, changes in delivery systems, or exposure to adverse events is fundamental to patient-centered outcomes research. Such analysis is critical for a wide range of comparative effectiveness research studies \citep{de2022trajectories, lycett2015association}, with examples including evaluating recovery trajectories after a post-fracture hospital-acquired condition, tracking HbA1c levels following the cessation of substance use, monitoring cognitive decline after initiating anti-aging medications, among others. Studying trajectories can offer an accurate representation of the actual patient experience, provide a clearer understanding of both the short- and long-term impacts of interventions or exposures, and uncover early or delayed effects that are critical for informed decision-making \citep{kern2013patient}. 

Real-world data, such as administrative claims and/or electronic health records, offer large sample sizes and provide a comprehensive understanding of how patients are influenced by specific exposures in real-world settings. These data enable more accurate and actionable insights into treatment efficacy, exposure effects, and patient experiences over time \citep{liu2022real}. Our case study focuses on evaluating the impact of hospital-acquired conditions (HAC) on post-discharge recovery trajectories in older adults living with Alzheimer’s Disease and Related Dementias (ADRD) who were hospitalized for hip fracture. Notably, older adults with ADRD are up to three times more likely than cognitively intact older adults to sustain a hip fracture \citep{taylor2013gait} and typically experience worse functional outcomes, greater disability, and increased dependency following a fracture \citep{livingston2020dementia}. Moreover, this high-risk subgroup is vulnerable to HAC, including infections or complications observed during hospitalization, which may further exacerbate their already poor post-operation outcomes \citep{gleason2015effect}. In this paper, we focus on a novel month-level post-discharge outcome called ``days spent at home" (DAH), a claims-based metric representing the average number of days patients %who are 
discharged alive spend at home. Greater DAH, reflecting lower utilization of institutional care, align with patient priorities for aging in place \citep{groff2016days}. Visualizing Medicare claims data reveals that 
\begin{figure}
    \centering
    \includegraphics[scale=0.44]{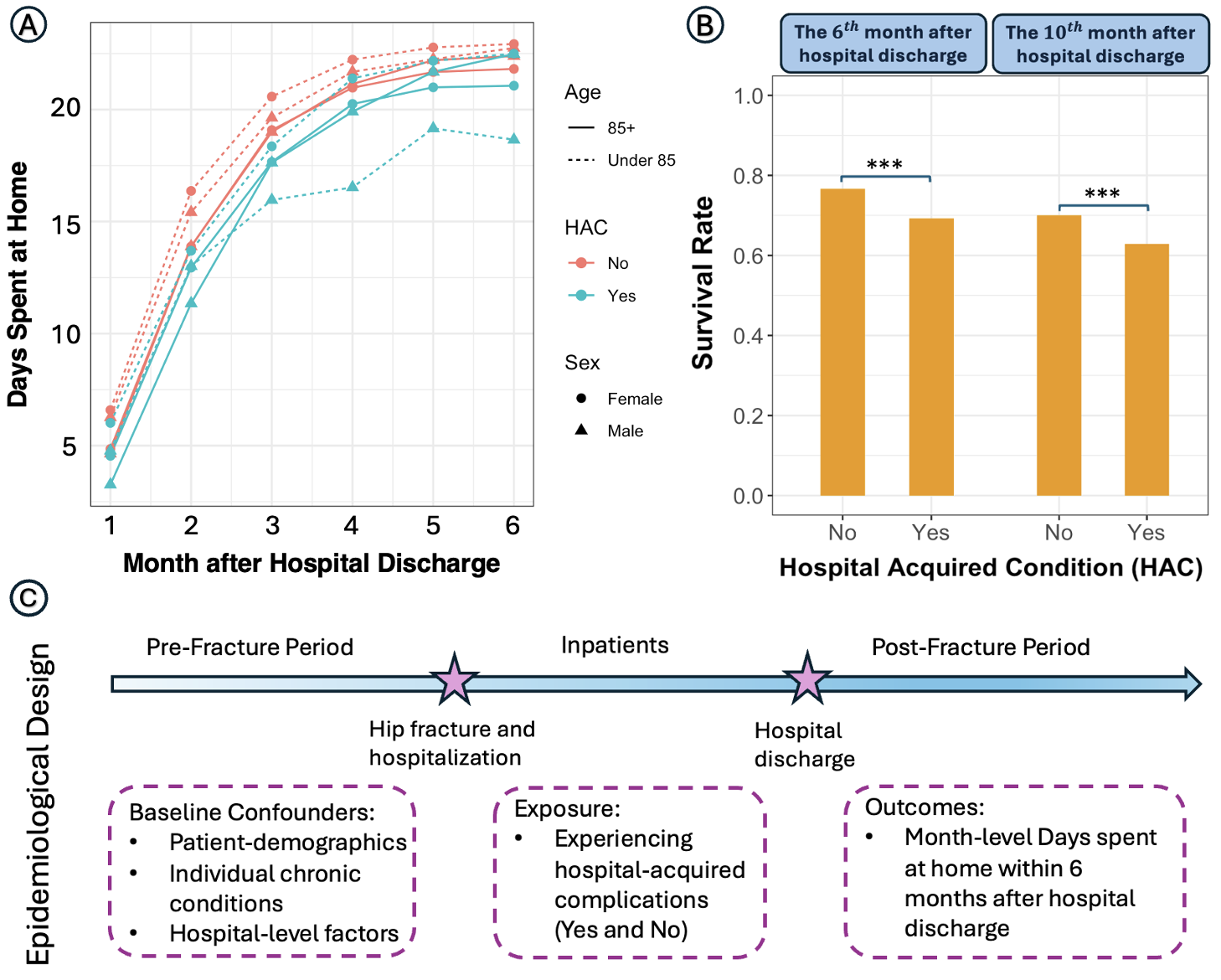}
    \caption{A. The DAH trajectories over six months after hospital discharge among older patients. B. The survival rates at month $6$ or $10$ after hospital discharge. Group comparisons were performed using two-sample proportion tests, with significance level $\ast\ast\ast: p\leq0.01$. C. Epidemiological design of data collection and question framing.}
    \label{introfigure}
\end{figure}
(\romannumeral 1) HAC may lead to fewer DAH, with effect sizes varying by month since discharge, possibly indicating delayed recovery; and (\romannumeral 2) HAC effects may vary substantially across sex and age categories (Figure \ref{introfigure}A). These observations strongly motivate the study of heterogeneity in HAC effects on DAH trajectories. We also observe a limited survival rate, $76.3\%$ at month $6$ after hospital discharge, with rates differing by HAC status: $69.3\%$ for patients with HACs and $76.7\%$ for those without (Figure \ref{introfigure}B). These significantly disproportionate survival rates suggest that mortality may be a critical intermediate factor that requires careful consideration when analyzing the heterogeneity of HAC effects. Relying solely on observed data from survivors is known to produce biased estimates \citep{rubin2006causal,ding2011identifiability}.
On the other hand, treating DAH as zero in the following months after death, although adopted in some clinical studies, is not appropriate for our setting. This approach fails to distinguish older adults with ADRD who die from those who survive but are subsequently admitted to skilled nursing facilities (SNFs) for long-term care. For ADRD patients following a fracture, discharge to SNFs is common and often clinically critical, as these facilities can provide specialized care \citep{abualrob2025scoping}. Collapsing both subgroups, death and prolonged SNF residence without death, into a single DAH value of zero distorts the observed outcome distribution, leads to misleading clinical interpretation of institutional care utilization, and consequently biases downstream analyses. Moreover, treating death merely as patient dropout is inappropriate, as outcomes cease to exist after death, which is fundamentally different from the mechanism of missingness \citep{rubin2006causal}. As an alternative, the principal stratification approach, detailed below, provides a promising framework to handle death.

\subsection{Methodology review}
To account for death as a post-exposure factor, principal stratification \citep{frangakis2002principal,tchetgen2014identification, ding2017principal,jiang2022multiply,lu2025principal} offers an appealing approach. It delineates causal mechanisms among patients who remain alive regardless of exposure levels, thus enabling valid causal interpretation of the effect within the principal stratum \citep{rubin2006causal,ding2011identifiability}. Principal stratification has also been successfully applied in a variety of applications, including non-compliance \citep{frumento2012evaluating,mealli2013using}, censored data \citep{frangakis1999addressing,mattei2014identification}, and surrogate outcomes \citep{frangakis2002principal,jiang2016principal}. It is worth highlighting that \cite{ding2017principal} proposed a novel identification and estimation approach under principal ignorability by introducing a principal score, which is defined as the conditional probability of the latent principal strata given covariates. In this context, the principal stratum refers to the sub-cohort of individuals who would remain alive regardless of exposure status. This approach is particularly useful when the assumptions required for classical instrumental variable analysis cannot be justified based on background knowledge or are incompatible with the scientific questions of interest \citep{ding2017principal}. Subsequently, \cite{jiang2022multiply} introduced a multiply robust approach for estimating average treatment effect (ATE) based on the principal ignorability assumption, which assumes that the latent principal strata are mean independent of the potential outcomes, conditional on the observed covariates. More recently, \cite{lu2025principal} developed nonparametric identification and semiparametric estimation theory for ATE-based principal stratification with continuous post-treatment variables. Beyond principal ignorability, the conditional independence of potential outcomes was also applied in identifying principal stratum estimand under survival analysis settings: for instance, \cite{lyu2023bayesian} studied survivor average causal effect (SACE) on recurrent events with a Bayesian framework, \cite{comment2025survivor} estimated the time-varing SACE for continuous time within the principal stratum. 

% However, existing principal stratification methods cannot be directly applied to infer HTE on longitudinal trajectories. More importantly, there has been little effort in the literature to characterize the patient profile of the principal stratum, a critical gap that must be addressed to enhance the broader applicability of principal stratification.

There are {several directions for further investigation in} principal stratification. (\romannumeral 1) Heterogeneity of treatment effects (HTE, the term ``treatment" can refer to either an exposure or an intervention) on an outcome trajectory in the presence of truncation by death. Existing methods primarily focus on average effects rather than effect modification \citep{egleston2009estimation,tchetgen2014identification, jiang2022multiply,lu2025principal,lyu2023bayesian,comment2025survivor,gonccalves2025survivor}. (\romannumeral 2) Characterization of the principal stratum. Owing to the latent nature, characterizing patient profiles within the unobserved principal stratum is challenging. Nevertheless, such characterization is crucial for improving the interpretation of the studied population. {(\romannumeral 3) Sensitivity analysis for assumption violation. Since identification of causal effects within principal strata typically relies on untestable assumptions, developing sensitivity analysis tailored for HTE analysis is essential for assessing the impact of potential departures from these assumptions.}

% The last challenge is well-recognized in the literature since real-world data is observational, meaning exposures, treatments, or interventions are not randomly assigned (Figure \ref{introfigure}C). Confounding can lead to systematic differences between exposed and unexposed groups, introducing bias \citep{liu2022real}. 

Before introducing our proposed method, we acknowledge that the past decade has seen significant advancements in statistical frameworks developed to infer HTE, also referred to as conditional average treatment effects (CATEs) to study effect modification \citep{kunzel2019metalearners}. These include, but are not limited to, tree-based approaches \citep{wager2018estimation}, pseudo outcome-based strategies \citep{semenova2021debiased}, {DR-learner \citep{kennedy2023towards} and R-learner \cite{nie2021quasi} focusing on infinite-dimensional estimation, weighted orthogonal learning approaches \citep{morzywolek2023weighted,shi2023meta} aiming at finite-dimensional projection-based parameter estimation through a working structural model. However, to the best of our knowledge, these methods cannot be directly applied to study effect modification in the presence of truncation by death. \cite{chen2024bayesian} proposed a Bayesian machine learning approach to estimate the heterogeneous survivor causal effects through Bayesian  nonparametric modeling of outcome regression and principal score. Their method targets the infinite-dimensional conditional treatment effect function and relies on Bayesian posterior inference. }

{We focus on a frequentist semi-parametric framework for studying HTE under truncation by death.} Two major contributions to existing literature are noteworthy. The first lies in its methodological innovation. {We propose a new estimand based on principal stratification, specifically tailored to the context of longitudinal trajectories where outcomes are truncated by death.} The new estimand is build upon projection techniques and an interpretable structural working model {focusing on finite dimensional parameter estimation \citep{kennedy2023semiparametric,ye2023instrumented,lu2025principal,zhang2025semiparametric}}, enabling the study of effect modification beyond the focus of average effects targeted by existing approaches under principal ignorability. We further establish identification for the working function and introduce {an efficient} \textbf{P}seudo \textbf{D}ata-based Triply \textbf{Robust} estimation procedure (PD-Robust). Importantly, PD-Robust framework also provides {model diagnosis and several sensitivity analysis strategies for evaluating the impact of assumption violations}. It further facilitates the characterization of patient profiles within the principal stratum, thereby making the principal stratum more interpretable in practice. The second contribution lies in the case study, where we revealed heterogeneity in HAC effects, with males under the age of $85$ years as a high-risk subgroup experiencing up to $23$ fewer days at home within six months post-discharge, comparing HAC to no HAC. This difference exceeds the 8-day
threshold regarded as clinically meaningful change in DAH due to any exposure over a six-month period \citep{auriemma2023stakeholder}. In addition, we have developed a user-friendly R package to facilitate real-world application of this method, with tutorial available at https://github.com/whhuan/PD\_Robust.

% from a methodologic prospective, it develops a triply robust and easily implementable causal inference framework for evaluating HTE on longitudinal trajectories in the presence of truncation by death, accompanied by detailed usage guidance to inform real data analysis; from an application prospective, it studies a novel claim-based post-fracture outcome for a high-risk sub-cohort by disentangling the heterogeneous HAC effects on post-fracture DAH trajectories for older adults with ADRD in a real-world setting, a topic that has been underexplored in existing studies. Specifically, 

% The case study 

The remainder of the paper is organized as follows: Section \ref{sec:meth} introduces the proposed method in a general setting. Section \ref{The case study} presents the case study with comprehensive statistical analyses. Section \ref{Numerical experiment} provides numerical evaluations through computer simulations.  Section \ref{Discussion} summarizes  and discusses the key findings. Technical proofs and extra numerical results are provided in the Supplementary Material.

\section{Methods}
\label{sec:meth}
\subsection{Notation and the proposed estimand}
%Introduce math notation, define estimand

% We consider a longitudinal data of $N$ subjects over $T$ months. For the $i$-th patient ($i=1,\ldots,N$) in the $t$-th month ($t=1,2,\ldots,T$), let $A_i\in\{0,1\}$ be the binary exposure (e.g, experiencing HAC during hospitalization), $\bfX_{i}$ be the vector of covariates serving as potential confounders (e.g., age, pre-facture DAH, chronic conditions, etc.), $Y_{i}(t)$ be the longitudinal outcome of interest (e.g., month-level post-fracture DAH after hospital discharge), $S_{i}(t)$ be the indicator of survival taking values $0$ if this subject is dead and $1$ if alive. To define our estimand of interest as causal and possibly interpret results as causal (per the assumptions detailed below), we adopt the counterfactual framework \citep{hernan2010causal}. Under the Stable Unit Treatment Value Assumption \citep{rubin1980discussion}, let $Y^a_{i}(t)$ and $S^a_{i}(t)$ be the potential values of the outcome and the survival status if the i-th patient were to receive the exposure $a$ with $a=0,1$, respectively.

We consider a data on $N$ patients over $T$ months. For the $i$-th patient ($i=1,\ldots,N$) in the $t$-th month ($t=1,2,\ldots,T$), let $\bfX_{i}$ be the vector of covariates serving as pre-fracture confounders (e.g., age, pre-fracture DAH, chronic conditions, etc.), $A_i=a\in\{0,1\}$ be the binary exposure at the baseline (e.g, experiencing HAC during hospitalization or not). To formally define our causal estimand and enable a causal interpretation of the results, we adopt the counterfactual framework \citep{hernan2010causal}. Under the Stable Unit Treatment Value Assumption \citep{rubin1980discussion}, if the $i$-th patient were to receive the exposure $a$ ($a=0,1$), let $Y^a_{i}(t)$ be potential value of the outcome of interest (e.g., post-fracture DAH during month $t$ after hospital discharge), and let $S^a_{i}(t)$ be potential survival status taking value $0$ if subject $i$ is dead and $1$ if alive. The observed outcome $Y_{i}(t)$ and survival indicator $S_{i}(t)$  correspond to the potential outcomes under the actual exposure $A_i$. 

To mitigate the challenges of survival bias in defining our causal estimand of interest, we focus on the causal effect for subjects who would survive until a pre-specified time point regardless of the exposure. This type of causal effect, often named principal causal effects (PCEs) \citep{frangakis2002principal}, accounts for the fact that death is an absorbing state \citep{rubin2006causal}, which differs from the mechanism of missing data in general (including missing potential outcomes), where data exist but are not observed. In this setting, the outcomes are fundamentally not defined for individuals who have died. Consequently, PCE may provide a more meaningful interpretation of causal effects in the presence of survival bias, which is a critical concern when analyzing data from older adults living with ADRD. Figure \ref{introfigure}B suggests potentially differential survival profiles between HAC statuses and may lead to biased HAC effect estimation if survival bias is not properly addressed. In addition to addressing survival bias, another challenging but intriguing task we aim to tackle is studying the HTE on an outcome trajectory between exposures $A=1$ and $A=0$, given $q$-dimensional covariates $\tilde{\bfX} \subset \bfX$, such as age and sex, in our case study. Below, we introduce a new estimand designed to achieve these goals, with the detailed rationale for proposing this estimand discussed in Section \ref{new perspectives}. 

In particular, let $t_0$ and $t_\ast$ be pre-specified timestamps satisfying $1\leq t_0< t_\ast\leq T$ and define $U(t)=\left(S^{a=1}(t),S^{a=0}(t)\right)\in\{0,1\}\times\{0,1\}$ as the joint potential survival status with exposures $A=1$ and $A=0$ at the $t$-th month ($t_0\leq t\leq t_\ast$). Following \cite{rubin2006causal}, the only reasonable sub-cohort of interest would be $U(t)=(1,1)$, i.e., survivors regardless of exposure status, for all $t\in[t_0,t_\ast]$. Thus, the new estimand for the causal trajectory from time $t_0$ to $t_\ast$ is defined {as a function of $t$ on the domain $[t_0,t_\ast]$:}  
\begin{equation}\label{new estimand}
    \tau_{(1,1)}(t,t_\ast,\tilde{\bfX})=\mathbb{E}\big(Y^{a=1}(t)-Y^{a=0}(t)|\tilde{\bfX},U(t_\ast)=(1,1)\big),1\leq t_0\leq t\leq t_\ast\leq T.
\end{equation}
 Briefly, suppose $\tilde{\bfX}=\tilde{\bfx}$ represents the profile for patients under $85$ years of age, for instance. In that case, the proposed estimand characterizes the average causal effect at time $t$ between exposure $A=1$ and $A=0$ for patients aged $85$ years old, who would survive at least until the $t_\ast$-th month, regardless of the exposure level.

In the following sections, we first explain the rationale behind the estimand in Equation (\ref{new estimand}) in Section \ref{new perspectives}. Then, we study its non-parametric identification based on the projection technique and propose a triply robust estimation procedure in Section \ref{The projection-based approximation} and \ref{The proposed estimation}. Last but not least, we provide practical guidance in Section \ref{Practical guidance} for implementing the method, including {model diagnosis, sensitivity analyses to assess the impact of assumption violations, and} strategies for characterizing patient profiles within the principal stratum.

\subsection{New perspectives on the proposed estimand}\label{new perspectives}
%yang + time
The proposed estimand in Equation (\ref{new estimand}) is unique in three key aspects: it (\romannumeral 1) incorporates a subset of covariates to enabling the assessment of effect modification, (\romannumeral 2) includes a time index $t$ to describe the outcome trajectory, and (\romannumeral 3) introduces a pre-specified timestamp $t_\ast$ for survival. The first two components intuitively extend existing PCE to an outcome trajectory and HTE. The third component is an additional feature that requires careful consideration. We provide the rationale for introducing this third component, focusing on the estimand without conditioning on $\tilde{\bfX}$ for illustration. The same rationale can be applied to the HTE setting.

In contrast to the estimand in Equation (\ref{new estimand}), one naive estimand %could be 
is $\tau_{(1,1)}(t)=\mathbb{E}\big(Y^{a=1}(t)-Y^{a=0}(t)|U(t)=(1,1)\big),t\in[1,T]$, i.e., the causal effect at time $t$ among patients who would survive at least until time $t$ regardless of  exposures $A=1$ and $A=0$. Although this estimand is interpretable at a given time $t$, the interpretation for the entire trajectory for $t \in [t_0, T]$ is less clear. The main reason is that the patient profiles of the principal stratum $U(t)=(1,1)$ can vary over time. {In particular, for any $t_1, t_2 \in [t_0, T]$ with $t_1<t_2$, if the patient belongs to either $U(t_1) = (0,1)$ or $U(t_1) = (0,0)$ at $t_1$, then this patient would not survive under one specific exposure at $t_1$, thus would not survive under that exposure at $t_2$ either. Therefore, the principal stratum $U(t_2)=(1,1)$ can only be a subset of $U(t_1)=(1,1)$ due to truncation by death.} Consequently, the patient profiles of these two strata can differ, and $\tau_{(1,1)}(t_1)$ and $\tau_{(1,1)}(t_2)$ do not consistently represent the average treatment effect on the same group of patients. This issue complicates estimation, interpretation, and reliable clinical decision-making.

% The joint potential survival status $U(t)$ defined in section 2.1 can serve as a principal stratification variable of the longitudinal data. For any individual $t\in[t_0,t_\ast]$, it is straightforward to define the estimand of the average treatment effect within the principal stratum $U(t)=s_as_{a'}$ as 
% $$\tau_{s_{\mathbf{a}}s_{\mathbf{a'}}}(t)=\mathbb{E}\big(Y_{\mathbf{a}}(t)-Y_{\mathbf{a'}}(t)|U(t)=s_{\mathbf{a}}s_{\mathbf{a'}}\big),s_as_{a'}\in\{00,01,10,11\}.$$
% Patients who would survive regardless of the treatments $A=a$ and $A=a'$ at this moment belong to the principal stratum $U(t)=11$, thus the local average treatment effect is 
% \begin{equation*}
% \tau_{s_{\mathbf{a}}s_{\mathbf{a'}}}(t)=\mathbb{E}\big(Y_{\mathbf{a}}(t)-Y_{\mathbf{a'}}(t)|U(t)=s_{\mathbf{a}}s_{\mathbf{a'}}\big),s_{\mathbf{a}}s_{\mathbf{a'}}=11,t_0\leq t\leq t_\ast\leq T.\tag{1}
% \end{equation*}

% \begin{equation*}
% \tau_{s_{\mathbf{a}}s_{\mathbf{a'}}}(t,t_\ast)=\mathbb{E}\big(Y_{\mathbf{a}}(t)-Y_{\mathbf{a'}}(t)|U(t_\ast)=s_{\mathbf{a}}s_{\mathbf{a'}}\big),s_{\mathbf{a}}s_{\mathbf{a'}}=11,t_0\leq t\leq t_\ast\leq T.\tag{2}
% \end{equation*}
% talk about interpretation of $ \tau_{s_{\mathbf{a}}s_{\mathbf{a'}}}(t,t',x)$ when $t=t'$, 
%highlight this estimand is valid for each individual t, however, the interpretation of the entire trajectory over different t will be tricky.
Our proposed estimand in Equation (\ref{new estimand}) addresses the interpretation issue. By introducing the additional survival timestamp $t_\ast$, patients who belong to the principal stratum $U(t_\ast)=(1,1)$ are consistent for the trajectory for $t\in[t_0,t_\ast]$. As a result, $\tau_{(1,1)}(t,t_\ast)$ represents the trajectory of causal effects on the sub-cohort of patients who would survive until the timestamp $t_\ast$. By varying the timestamp $t_\ast$, multiple causal effects $\tau_{(1,1)}(t,t_\ast)$ may arise and the difference between the areas under the curve (AUC) of these causal effects may be used to describe the potential population shifts over time. We formally define and interpret this metric in Section $2$ of the Supplementary Materials.

\subsection{The projection-based approximation}\label{The projection-based approximation}
% describe the formula here and provide rationale why it is triply robust and why it is more preferred than other estimation

% As illustrated in Section 1 of the Supplementary Material, to unbiasedly estimate (\ref{new estimand}), one may need to correctly specify the joint data distribution  conditional on $\tilde{\bfX}$. However, 

Despite its advantages, the estimand defined in Equation (\ref{new estimand}) may represent a complex functional of $\tilde{\bfX}$, making it difficult to identify and estimate \citep{geenens2011curse}. It may not be easily interpretable in terms of effect modification sizes and their statistical significance.  
% Notably, even under the case where the full data distribution conditional on all the covariates $\bfX$ follows a simple linear structure, the conditional distribution based on the reduced covariate set $\tilde\bfX$ can become complex and nonlinear after marginalization. 
% non-parametrically estimating these quantities may be infeasible in practice due to the curse of dimensionality when there are more than three effect modifiers under consideration \citep{geenens2011curse}.
Therefore, 
% model misspecification for the HTE estimation may be unavoidable in practice, making 
it is crucial to adopt a structural, working, interpretable model that reasonably approximates the estimand $\tau_{(1,1)}(t,t_\ast,\tilde{\bfX})$. In what follows, we define a structural working function based on a first moment constraint and describe how this function can be identified. A triply robust estimation procedure is then presented in Section \ref{The proposed estimation}.

{Let $f(t,t_\ast,\tilde\bfX;\bfbeta_t)$ be a structural working function indexed by a parameter vector $\bfbeta_t$ (which may also vary over time $t_\ast$; the subscript is omitted for brevity), of which the underlying value $\bfbeta_{t0}$ is defined as
\begin{equation}\label{wt_lsq_loss}
\bfbeta_{t0}=\arg\min_{\bfbeta_t}\mathbb{E}\Big[ \big\{\tau_{(1,1)}(t,t_\ast,\tilde\bfX)-f(t,t_\ast,\tilde\bfX;\bfbeta_{t})\big\}^2\Big|U(t_\ast)=(1,1)\Big].
\end{equation}
Intuitively, the structural working function $f(t,t_\ast,\tilde\bfX;\bfbeta_{t0})$ can be interpreted as a projection of $\tau_{(1,1)}(t,t_\ast,\tilde\bfX)$ onto the space defined by the working model $f(t,t_\ast,\tilde\bfX;\bfbeta_t)$, thus representing a reasonable approximation of the underlying quantity. The parameters $\bfbeta_{t0}$ is then the solution to the following estimating equation
\begin{equation}\label{def of working fct}
    \mathbb{E}\Big[\bfg(t,t_\ast,\tilde\bfX;\bfbeta_{t0})
    %\mathbb{P}(U(t_\ast)=(1,1)|\tilde{\bfX})
    \big\{\tau_{(1,1)}(t,t_\ast,\tilde\bfX)-f(t,t_\ast,\tilde\bfX;\bfbeta_{t0})\big\}\Big|U(t_\ast)=(1,1)\Big]=\bfzero
\end{equation}
by taking the first order derivative with respect to $\bfbeta_{t0}$ for the formulation in (\ref{wt_lsq_loss}), where $\bfg(t,t_\ast,\tilde\bfX;\bfbeta_{t0})=\partial f(t,t_\ast,\tilde\bfX;\bfbeta_{t0})/\partial \bfbeta_{t0}$.
% Intuitively, \textcolor{red}{the structural working function $f(t,t_\ast,\tilde\bfX;\bfbeta_{t0})$ can be interpreted as a projection of $\tau_{(1,1)}(t,t_\ast,\tilde\bfX)$ onto the space
% defined by the working model $f(t,t_\ast,\tilde\bfX;\bfbeta_t)$}. This can be justified by acknowledging the fact that the parameters $\bfbeta_{t0}$ solving Equation (\ref{def of working fct}) generally are the minimizers of the weighted least-square loss function $\mathbb{E}\Big[\omega(t,t_\ast,\tilde\bfX) 
% %\mathbb{P}(U(t_\ast)=11|\tilde{\bfX}) 
% \big\{\tau_{(1,1)}(t,t_\ast,\tilde\bfX)-f(t,t_\ast,\tilde\bfX;\bfbeta_{t0})\big\}^2\Big|U(t_\ast)=(1,1)\Big]$ with some weight function $\omega(t,t_\ast,\tilde\bfX)$, thus representing a reasonable approximation of the underlying quantity.
Moreover, given the conditional feature in equation (\ref{def of working fct}), the projection is made only based on the principal stratum $U(t_\ast)=(1,1)$, which is reasonable as only subjects belonging to the principal stratum are of interest.
    % The term $\mathbb{P}(U(t_\ast)=11|\tilde{\bfX})$ represents the true conditional probability of $U(t_\ast)=11$ and appears in the above moment condition due to certain technical reasons. However, as we show in Section \ref{The proposed estimation}, there is no need to specify $\mathbb{P}(U(t_\ast)=11|\tilde{\bfX})$ in order to achieve a robust and unbiased estimate of $\bfbeta$. 
    In practice, end-users can flexibly specify the structural working function to meet specific needs. One typical example is to specify $f(t,t_\ast,\tilde\bfX;\bfbeta_t)=\bfeta^T(\tilde\bfX)\bfbeta_t$ and $\bfg(t,t_\ast,\tilde\bfX;\bfbeta_t)=\bfeta(\tilde\bfX)$ when the outcome is continuous, and $f(t,t_\ast,\tilde\bfX;\bfbeta_t)=l(\bfeta^T(\tilde\bfX)\bfbeta_t)$ with some link function $l(\cdot)$ when the outcome is categorical. We considered the linear form in Section \ref{The case study} and Section \ref{Numerical experiment}.}

It is important to highlight that the proposed structural working function is not directly computable as it involves {unobserved latent variables}. Additional assumptions are therefore required to identify the underlying parameter vector $\bfbeta_{t0}$. We introduce three nuisance functions and several assumptions to establish the identification of $\bfbeta_{t0}$ in Equation (\ref{def of working fct}):

\begin{enumerate}
    \item The propensity score $\pi(\bfX)$ (PS): the probability of being exposed to $A=1$ conditional on the covariates $\bfX$, i.e., $\pi(\bfX)=\mathbb{P}(A=1|\bfX)$; 
    \item The {principal score} $e_{(s_1,s_0)}(t_\ast,\bfX)$ (PPS): the probability of being in principal stratum $U(t_\ast)=\left(s_1,s_0\right)$, where $s_1,s_0\in\{0,1\}$, at time $t_\ast$ conditional on the covariates $\bfX$ and the exposures $A=1$ and $A=0$, i.e., $e_{(s_1,s_0)}(t_\ast,\bfX)=\mathbb{P}\left(U(t_\ast)=\left(s_1,s_0\right)|\bfX\right)$. We further denote $p_a(t_\ast,\bfX)=\mathbb{P}(S(t_\ast)=1|A=a,\bfX)$ as the probability of observing survival conditional on the exposure and covariates; 
    \item The {conditional outcome mean} $\mu_{as}(t,t_\ast,\bfX)$ (CM): the mean of the outcome at time $t$ conditional on the treatment, the observed survival status at time $t_\ast$, and covariates, i.e., $\mu_{as}(t,t_\ast,\bfX)=\mathbb{E}(Y(t)|A=a,S(t_\ast)=s,\bfX)$ for $a,s\in\{0,1\}$.
\end{enumerate}

These three nuisance functions described above are often unknown when analyzing real-world data. Confounding may exist between the exposure and survival status, as well as between the exposure and the outcome, and should be appropriately accounted for using the observed data. Moreover, the value of $U_i(t_\ast)$ cannot be observed directly since we cannot observe $S^{a=1}_{i}(t_\ast)$ and $S^{a=0}_{i}(t_\ast)$ simultaneously for any $i=1,\ldots,N$. We detail how these functions can be estimated in Section \ref{The proposed estimation}. 

We now introduce four commonly adopted assumptions in studying PCE \citep{ding2017principal,jiang2022multiply,lu2025principal} to facilitate the identification of $\bfbeta_{t0}$: 

\noindent\textbf{Assumption} $\mathbf{1}$ (Positivity). $0<\pi(\bfX)<1$.

\noindent\textbf{Assumption} $\mathbf{2}$ (Exposure ignorability). For $t\in[t_0, t_\ast]$ and $a\in\{0,1\}$, {$A\perp\mkern-10mu\perp(S^a(t),Y^a(t))|\bfX$.}

\noindent\textbf{Assumption} $\mathbf{3}$ (Monotonicity). For $i=1,\ldots,N$ {and any $t\in[t_0,t_\ast]$, $S^{a=1}_{i}(t)\leq S^{a=0}_{i}(t)$.}

\noindent\textbf{Assumption} $\mathbf{4}$ (Principal ignorability). $\mathbb{E}\big(Y^{a=1}(t)|U(t_\ast)=(0,0),\bfX\big)=\mathbb{E}\big(Y^{a=1}(t)|U(t_\ast)\\=(0,1),\bfX\big)$ and $\mathbb{E}\big(Y^{a=0}(t)|U(t_\ast)=(1,1),\bfX\big)=\mathbb{E}\big(Y^{a=0}(t)|U(t_\ast)=(0,1),\bfX\big)$. 

Assumption $1$ is reasonable as each patient will have some chance of experiencing HAC. Assumption $2$ states that, given the covariates, there are no unmeasured confounders of the actual exposure level and the counterfactual survival status and longitudinal outcomes during the time period $[t_0, t_\ast]$. It is important to note that potential outcomes and survival status is allowed to be associated even after adjusting for all covariates. This assumption provides a more realistic framework for handling death when it is intrinsically linked to the outcome. {In our data application, we included patient demographics, clinical conditions, and hospital-level factors to mitigate the unmeasured confounding issue. In Section S$1.6.2$ of the Supplementary Material, we provided a sensitivity analysis strategy for assessing the impact of violations of this assumption.} Assumption $3$ requires that the non-exposure $A = 0$ has a non-negative impact on survival status compared to the exposure $A = 1$ for each patient at $t_\ast$, thereby ruling out the stratum $U(t_\ast) = (1,0)$. This %assumption 
is reasonable in our case study, as HAC is expected to adversely affect survival. Also, Assumptions $2$ and $3$ make the principal score estimable: $e_{(1,1)}(t_\ast,\bfX)=p_1(t_\ast,\bfX)$. Assumption $4$ holds if the conditional expectations of the potential outcomes are influenced only by the survival status evaluated at the same exposure level. For instance, the survival status under $a=1$ and covariates $\bfX$ sufficiently explain the mean of the potential outcome under $a=1$; thus, this mean is no longer associated with the survival status under $a=0$ after controlling for the survival status under $a=1$ and covariates $\bfX$. This assumption may hold in our case study, where DAH has shown to be highly associated with mortality among ADRD older adults after experiencing a hip fracture \citep{shen5378827joint}. {In Section \ref{Practical guidance}, we describe a sensitivity analysis strategy for assessing the impact of violations of this assumption.} The following lemma ensures the non-parametric identification of $\bfbeta_{t0}$.

\begin{lemma}\label{lemma 1}
Suppose that Assumptions $1$-$4$ hold and $e_{(1,1)}(t_\ast,\bfX) > 0$ for all $\bfX$. Then there exists an identification formula for $\bfbeta_{t0}$, based on any two combinations of likelihood components: $\pi(\bfX)$, $e_{(1,1)}(t_\ast,\bfX)$ and $\mu_{as}(t,t_\ast,\bfX)$. 
\end{lemma}

As discussed in Section S$1.1$ and Section S$1.3$ of the Supplementary Material, the above identification of $\bfbeta_{t0}$ leads to several possible estimators, each requiring the correct specification of two components of the observed distribution. 
% Inspired by the work of \cite{jiang2022multiply} where average effects were studied, this feature motivates the formulation of a more robust estimator to study HTE, which is illustrated in the next Section. 

\subsection{A Pseudo Data-based Robust estimation (PD-Robust)}\label{The proposed estimation}

{Since Lemma \ref{lemma 1} establishes that multiple estimators of the HTE can be constructed, an natural question is which estimator is robust and efficient. This consideration motivates the derivation of the efficient influence function (EIF) for the projection parameter $\bfbeta_{t0}$ at each time $t\in[t_0,t_\ast]$ when the working model has a linear form, i.e., $f(t,t_\ast,\bfX;\bfbeta_t)=\bfeta^T(\bfX)\bfbeta_t$. As detailed in Section S$1.5$ of the Supplementary Material, the resulting EIF form provides the basis for the proposed estimation procedure (PD-Robust) described below.
% , which is obtained by solving an estimating equation targeting the working model defined in Equation (\ref{def of working fct}). Specifically, if the working model has a linear form, i.e., $f(t,t_\ast,\bfX;\bfbeta_t)=\bfeta^T(\bfX)\bfbeta_t$, then the EIF is also linear in $\bfbeta_t$, thus the solution to the estimating equation will have a closed form. We introduce the detailed derivation of the EIF in Section S$1$ of the Supplementary material.
% thus we propose the proced$ure of PD-Robust (Figure \ref{workflow}) to yield an estimator achieving semiparametric efficiency bound when all three nuisance models are correctly specified. 
Moreover, such estimator is also triply robust and normally distributed in large samples.} An estimator is triply robust in the sense that the estimated parameters, denoted by $\hat\bfbeta_t$, will be a consistent estimator if any two of these three nuisance functions are correctly modeled. We begin by presenting the PD-Robust estimator directly, followed by more explanation of the intuition behind the procedure.
Specifically, given $t\in[t_0,t_\ast]$, let $\pi(\bfX;\bfalpha)$, $e_{(1,1)}(t_\ast,\bfX;\bftheta)$, and $\mu_{as}(t,t_\ast,\bfX;\bfgamma)$ be the working models for $\pi(\bfX)$, $e_{(1,1)}(t_\ast,\bfX)$, and $\mu_{as}(t,t_\ast,\bfX)$, respectively. The estimation of the nuisance parameters $\hat{\bfalpha}$, $\hat{\bftheta}$, and $\hat{\bfgamma}$ in these models can be obtained using parametric or machine learning methods (\cite{semenova2021debiased}). Once all nuisance models are obtained, the PD-Robust procedure for estimating $\bfbeta_{t0}$ can be achieved by following the three steps to create pseudo data: 

\textbf{Step 1}: Obtain estimates of parameters $\hat{\bfalpha}$, $\hat{\bftheta}$, and $\hat{\bfgamma}$ from nuisance models;

\textbf{Step 2}: For each subject $i$ and a given $t_\ast$, calculate the pseudo outcome $\hat{\phi}_{1,i}-\hat{\phi}_{0,i}$ and the pseudo scalar $\hat{\psi}_{S_0,i}$ at each time $t$ based on the estimated nuisance models $\pi(\bfX_i;\hat{\bfalpha})$, $e_{(1,1)}(t_\ast,\bfX_i;\hat{\bftheta})$, and $\mu_{as}(t,t_\ast,\bfX_i;\hat{\bfgamma})$, where
\begin{equation*}
\begin{split}
\hat{\phi}_{1,i}&=\frac{\mathbf{1}\{A_i=1\}[Y_i(t)S_i(t_\ast)-\mu_{11}(t,t_\ast,\bfX_i;\hat{\bfgamma})p_1(t_\ast,\bfX_i;\hat{\bftheta})]}{\pi(\bfX_i;\hat{\bfalpha})}+\mu_{11}(t,t_\ast,\bfX_i;\hat{\bfgamma})p_1(t_\ast,\bfX_i;\hat{\bftheta}),\\
    \hat{\phi}_{0,i}&=\frac{e_{(1,1)}(t_\ast,\bfX_i;\hat{\bftheta})S_i(t_\ast)\mathbf{1}\{A_i=0\}}{p_0(t_\ast,\bfX_i;\hat{\bftheta})[1-\pi(\bfX_i;\hat{\bfalpha})]}[Y_i(t)-\mu_{01}(t,t_\ast,\bfX_i;\hat{\bfgamma})]+\mu_{01}(t,t_\ast,\bfX_i;\hat{\bfgamma})\hat{\psi}_{S_1,i},\\
    \hat{\psi}_{S_1,i}&=\frac{\mathbf{1}\{A_i=1\}[S_i(t_\ast)-p_1(t_\ast,\bfX_i;\hat{\bftheta})]}{\pi(\bfX_i;\hat{\bfalpha})}+p_1(t_\ast,\bfX_i;\hat{\bftheta}).
    \end{split}
\end{equation*}
% \begin{equation*}
% \begin{split}
% \hat{\phi}_{1,i}&=\frac{\mathbf{1}\{A_i=1\}[Y_iS_i-\mu_{11}(\bfX_i;\hat{\bfgamma})p_1(\bfX_i;\hat{\bftheta})]}{\pi(\bfX_i;\hat{\bfalpha})}+\mu_{11}(\bfX_i;\hat{\bfgamma})p_1(\bfX_i;\hat{\bftheta}), (1)\\
%     \hat{\phi}_{0,i}&=\frac{e_{(1,1)}(\bfX_i;\hat{\bftheta})S_i\mathbf{1}\{A_i=0\}}{p_0(\bfX_i;\hat{\bftheta})[1-\pi(\bfX_i;\hat{\bfalpha})]}[Y_i-\mu_{01}(\bfX_i;\hat{\bfgamma})]+\mu_{01}(\bfX_i;\hat{\bfgamma})\hat{\psi}_{S_1,i},(2)\\
%     \hat{\psi}_{S_1,i}&=\frac{\mathbf{1}\{A_i=1\}[S_i-p_1(\bfX_i;\hat{\bftheta})]}{\pi(\bfX_i;\hat{\bfalpha})}+p_1(\bfX_i;\hat{\bftheta}).(3)\\
%     &\tilde\bfX\{ \hat{\phi}_{1,i}-\hat{\phi}_{0,i}-\hat{\psi}_{S_1,i}\tilde\bfX\bfbeta\}.(4)
%     \end{split}
% \end{equation*}

\textbf{Step 3}: Obtain $\hat{\bfbeta}_t$ by solving the estimating equation using pseudo outcome and scalar:
\begin{equation}\label{estimating.eq}
\sum_{i=1}^N\bfg(t,t_\ast,\tilde\bfX_i;\bfbeta_t)\{ \hat{\phi}_{1,i}-\hat{\phi}_{0,i}-\hat{\psi}_{S_1,i}f(t,t_\ast,\tilde\bfX_i;\bfbeta_t)\}=\bfzero.
\end{equation}

There are Two key features of PD-Robust are worth highlighting. 
% First, the above estimation procedure acknowledges that the underlying principal stratum label $U(t_\ast)=(1,1)$ is unknown.
% , even though it appears in the definition of the working function in (\ref{def of working fct}). 
First, PD-Robust is computationally efficient, given that all pseudo data can be conveniently calculated without complex integration. When a linear model is proposed for $f(\cdot)$, PD-Robust will lead to a closed-form estimator. Second, the resulting estimator $\hat\bfbeta_t$ is consistent, if any two of the three working models for $\pi(\bfX), e_{(1,1)}(t_\ast,\bfX),\text{ and }\mu_{as}(t,t_\ast,\bfX)$ are correctly specified. Intuitively, this property holds due to key observations: $\mathbb{E}({\phi}_{1}-{\phi}_{0})=\mathbb{E}[\{Y^{a=1}(t)-Y^{a=0}(t)\}I_{\{U(t_\ast)=(1,1)\}}]$, and
$\mathbb{E}({\psi}_{S_1})=\mathbb{P}(U(t_\ast)=(1,1))$, provided that at least one pair among $\big\{\pi(\bfX;\bfalpha), e_{(1,1)}(t_\ast,\bfX;\bftheta)\big\}$, $\big\{e_{(1,1)}(t_\ast,\bfX;\bftheta), \mu_{as}(t,t_\ast,\bfX;\bfgamma)\big\}$ or $\big\{\pi(\bfX;\bfalpha),\mu_{as}(t,t_\ast,\bfX;\bfgamma)\big\}$ equals the true nuisance functions. It is not necessary for all pairs to be correct simultaneously. %Thus, 
Under this condition, we can show that $\mathbb{E}[\bfg(t,t_\ast,\tilde\bfX;\bfbeta_{t0})\{ {\phi}_{1}-{\phi}_{0}-{\psi}_{S_1}f(t,t_\ast,\tilde\bfX;\bfbeta_{t0})\}]=\bfzero$, which ensures a consistent estimator \citep{van2000asymptotic}. 
% Let $\mathcal{M}_{tp+ps}$, $\mathcal{M}_{tp+om}$, and $\mathcal{M}_{ps+om}$ represent the pairs of correctly specified models $\{\pi(X), e_{11}(t,X)\}$, $\{\pi(X), \mu_{zs}(t,X)\}$ and $\{e_{11}(t,X),\mu_{zs}(t,X)\}$, respectively.
We summarize in Theorem \ref{thm1} the triple robustness and the asymptotic property of $\hat{\bfbeta}_t$, with detailed proof provided in Section S$1.2$ of the Supplementary Material.

\begin{figure}
    \centering
    \includegraphics[scale=0.35]{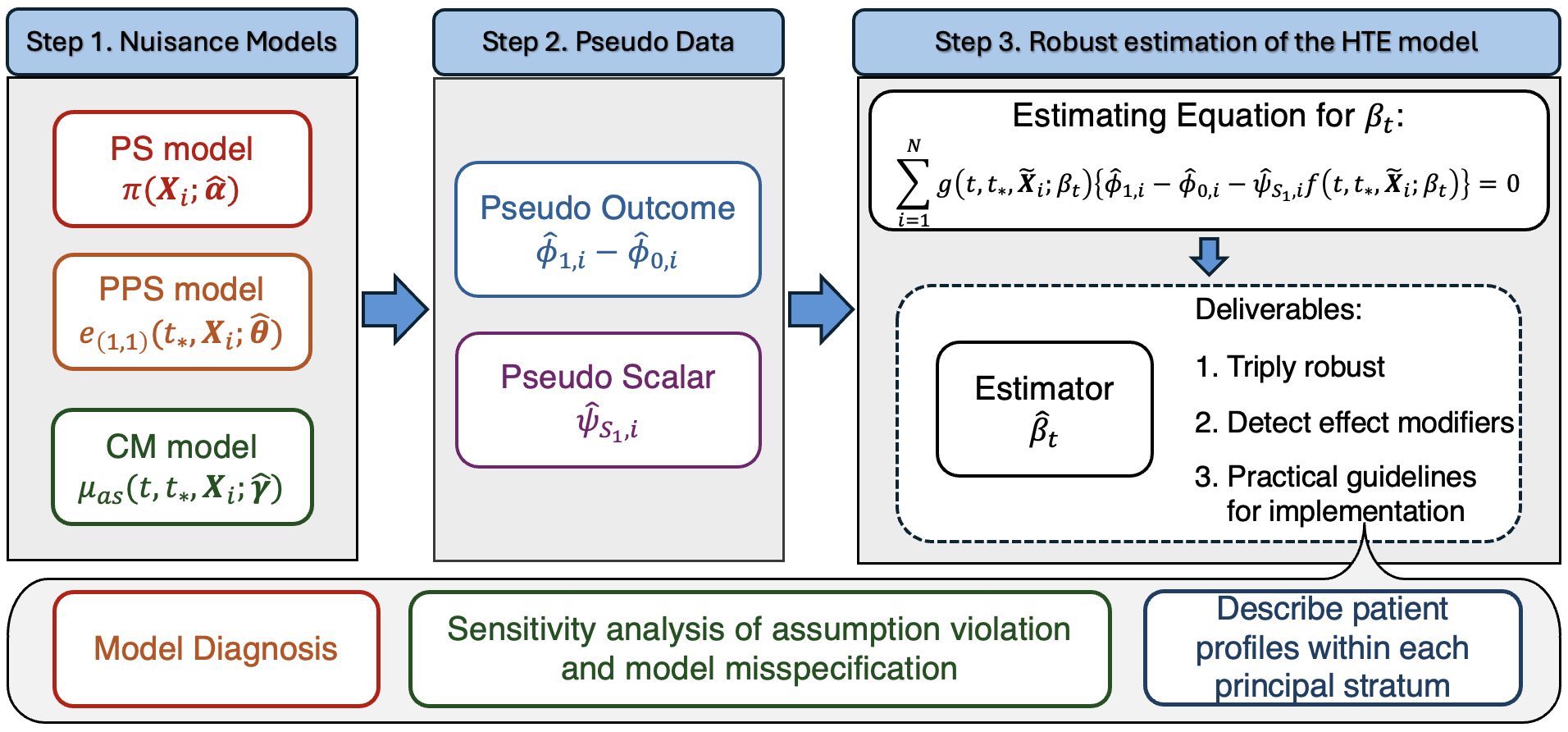}
    \caption{{The practice guideline of the PD-Robust in obtaining the HTE parameter estimate, evaluating model specification and assumption validation, and characterizing patient profile within the principal strata.}}
    \label{workflow}
\end{figure}

\begin{theorem}\label{thm1}
Suppose that Assumptions $1$-$4$ hold. Under extra conditions listed in the Supplementary Material, the estimator $\hat\bfbeta_t$ obtained by solving Equation (\ref{estimating.eq}) is triply robust in the sense that it converges in probability to $\bfbeta_{t0}$ defined in Equation (\ref{def of working fct}) if any two of the three working models for PS, PPS, and CM are correctly specified. Moreover, the estimator $\hat\bfbeta_t$ follows an asymptotically normal distribution with expectation $\bfbeta_{t0}$.
\end{theorem}

%, with mean zero and variance formula displayed in the Supplementary Material.
% {yilin: the proof for theorem 4 and 5 from ding peng's paper (Principal Stratification with Continuous
% Post-Treatment Variables: Nonparametric
% Identification and Semiparametric Estimation) may be more relevant to yours since they are focusing on Z-estimator, while yang shu and debiased paper focus on a closed form estimator, which is a bit different from your case. i will guide you go over the proof procedure when we are there .} 

Note that the derived asymptotic variance in Section S$1.2$ of the Supplementary Material depends on the correct specification of nuisance functions, and in practice, the bootstrap procedure is recommended for statistical inference. Moreover, the multiplicative nature of the two errors in PD-Robust estimation also enables the use of machine learning to estimate nuisance functions, aligning with conventional causal machine learning approaches \citep{chernozhukov2018double, semenova2021debiased}. For further theoretical discussions, we refer interested readers to Section S$2$ of the Supplementary Material.

\subsection{Practical guidance}\label{Practical guidance}
% In this section, we will talk about model diagnosis, including PS, PPS, and sensitivity analyses for Mu, also how to identify patients associated with the principal stratum, which will align better with our real data application. no need to be very technically, and try to use language to deliver information.  

In this section, we provide several guidelines (Figure \ref{workflow}) to enhance the applicability of PD-Robust, including: (\romannumeral 1) {diagnosis for the PS and PPS models,  (\romannumeral 2) violation of idenfication assumptions, particularly the principal ignorability assumption;} (\romannumeral 3) identifying patient profiles within each of the {principal strata $U(t_\ast)=(1,1)$, $(0,1)$ and $(0,0)$.} 
% A detailed example of model checking and principal stratum identification using these guidelines is illustrated in the case study in Section \ref{The case study}.

\textbf{PS and PPS models.} A common strategy to evaluate how well a PS model performs is to evaluate if the distributions of baseline covariates are well balanced between groups with and without exposure following PS adjustment.  Specifically, we calculate the standardized mean difference (SMD) for each covariate using PS weighted
samples. Following established guidelines \citep{zhang2019balance}, an SMD value below $0.1$
indicates adequate balance. On the other hand,  motivated by the theoretical result that 
% the distribution of baseline covariates should be the same in both the group with and without the exposure conditional on both PS and PPS (\cite{jiang2022multiply}), 
the expectation of baseline covariates weighted by the inverse of PS and PPS should be identical between the exposure groups when the PPS model is correctly specified \citep{jiang2022multiply}, i.e.,
$$\mathbb{E}\left\{\frac{p_0(t_\ast,\bfX)}{p_0(t_\ast)}\frac{S(t_\ast)}{p_1(t_\ast,\bfX)}\frac{A}{\pi(\bfX)}\bfX\right\}=\mathbb{E}\left[\frac{S(t_\ast)(1-A)}{p_0(t_\ast)\{1-\pi(\bfX)\})}\bfX\right],$$
we recommend assessing the plausibility of the PPS model using this %balance 
property. Specifically, we compute the standardized $t$-statistic for the PCE across all baseline covariates, where a small absolute value indicates no PCE on baseline covariates as expected. Models %that 
achieving satisfactory balance suggest appropriate specification of the PS and PPS models, although balance alone does not guarantee correct model specification. Note that complex machine learning algorithms do not necessarily yield better causal estimates \citep{chen2025effect} in a finite sample setting and typically requires all estimated nuisance models to be close to their true functions %in order
to enable tractable inference \citep{chernozhukov2018double, jiang2022multiply}. %Thus, 
In practice, parametric models that achieve satisfactory balance may be preferable.

{\textbf{Sensitivity analysis for violations of the principal ignorability assumption.} Unlike monotonicity assumption, which is plausible in our applications because HAC is generally believed to worsen patients' survival outcomes, the validity of the principal ignorability assumption is more difficult to assess and cannot be directly verified from the observed data. To this end, we extend the existing sensitivity analysis strategy \citep{cheng2026multiply,jiang2022multiply} to investigate the robustness of our proposed HTE estimator within the $U(t_\ast)=11$ stratum against the violation of Assumption $4$, i.e., the following ratio
\begin{equation*}
    % \epsilon_1(t,t_\ast,\bfX)=\frac{\mathbb{E}\left(Y^{a=1}(t)|U(t_\ast)=(0,1),\bfX\right)}{\mathbb{E}\left(Y^{a=1}(t)|U(t_\ast)=(0,0),\bfX\right)},
    \epsilon_0(t,t_\ast,\bfX)=\frac{\mathbb{E}\left(Y^{a=0}(t)|U(t_\ast)=(0,1),\bfX\right)}{\mathbb{E}\left(Y^{a=0}(t)|U(t_\ast)=(1,1),\bfX\right)}
\end{equation*}
does not equal $1$. Specifically, while the pseudo outcome $\hat{\phi}_{1,i}$ and the pseudo scalar $\hat{\psi}_{S_1,i}$ remain unaffected, the pseudo outcome $\hat{\phi}_{0,i}$ can be formulated by incorporating $\epsilon_0(t,t_\ast,\bfX)$ under Assumption $1$-$3$ as
\begin{equation*}
\begin{split}
    \hat{\phi}_{0,i}&=\frac{w_0(t,t_\ast,\bfX)e_{(1,1)}(t_\ast,\bfX_i;\hat{\bftheta})}{p_0(t_\ast,\bfX_i;\hat{\bftheta})}\bigg[\frac{\mathbf{1}\{A_i=0\}[Y_i(t)S_i(t_\ast)-\mu_{01}(t,t_\ast,\bfX_i;\hat{\bfgamma})p_0(t_\ast,\bfX_i;\hat{\bftheta})]}{1-\pi(\bfX_i;\hat{\bfalpha})}\\
    &+\mu_{01}(t,t_\ast,\bfX_i;\hat{\bfgamma})p_0(t_\ast,\bfX_i;\hat{\bftheta})\bigg]-\frac{w_0^2(t,t_\ast,\bfX)\mu_{01}(t,t_\ast,\bfX_i;\hat{\bfgamma})}{\epsilon_0(t,t_\ast,\bfX)}\frac{p_1(t_\ast,\bfX_i;\hat{\bftheta})}{p_0(t_\ast,\bfX_i;\hat{\bftheta})}\\
    &\bigg[\frac{\mathbf{1}\{A_i=0\}[S(t_\ast)-p_0(t_\ast,\bfX_i;\hat{\bftheta})]}{1-\pi(\bfX_i;\hat{\bfalpha})}+p_0(t_\ast,\bfX_i;\hat{\bftheta})\bigg]
    +\frac{w_0^2(t,t_\ast,\bfX)\mu_{01}(t,t_\ast,\bfX_i;\hat{\bfgamma})\hat{\psi}_{S_1,i}}{\epsilon_0(t,t_\ast,\bfX)}\\
    \end{split}
\end{equation*}
with $w_0(t,t_\ast,\bfX)=p_0(t_\ast,\bfX_i;\hat{\bftheta})/[\epsilon_0(t,t_\ast,\bfX)(p_0(t_\ast,\bfX_i;\hat{\bftheta})-p_1(t_\ast,\bfX_i;\hat{\bftheta}))+p_1(t_\ast,\bfX_i;\hat{\bftheta})]$.
The detailed derivation can be found in Section S$1.5$ of the Supplementary Material. Since the patients in the stratum $U(t_\ast)=(1,1)$ would survive at least till time $t_\ast$ regardless of the exposure, it is reasonable to assume that the patients in this stratum are healthier than patients in other strata, thus $\epsilon_0(t,t_\ast,\bfX)\leq 1$. Therefore, we can set $\epsilon_0(t,t_\ast,\bfX)$ to vary within $(0,1]$ to evaluate how sensitive the estimator is to varying extent of the violation of the principal ignorability assumption.}

{Moreover, we provide a latent-variable-based sensitivity analysis framework to evaluate the impact of violations of the exposure ignorability assumption. Details are provided in Section S$1.6.2$ of the Supplementary Material.}

\textbf{Characterization of patient profiles associated with the principal stratum.} As described in Section \ref{sec:meth}, we focus on the patients who would survive regardless of exposure for all $t\in[t_0,t_\ast]$, 
% Since the patients who survive at $t_2$ must survive at $t_1$ for any $t_1,t_2$ such that $t_0\leq t_1<t_2\leq t_\ast$, the principal stratum of interest consists of patients who would survive at $t_\ast$ regardless of the exposure, 
i.e., those satisfying $U(t_\ast)=(1,1)$, which is equivalent to $S^{a=1}(t_\ast)=1$ by Assumptions $2$ and $3$. Under {these two assumptions}, this principal stratum can be further expressed as the observed stratum $\{S(t_\ast)=1|A=1\}$. Thus, given covariates $\bfX$, the PPS $e_{(1,1)}(t_\ast,\bfX)$ is identified as $e_{(1,1)}(t_\ast,\bfX)=\mathbb{P}\left(S(t_\ast)=1|A=1,\bfX\right)$,
which can be approximated using a working model $p_1(t_\ast,\bfX;\boldsymbol{\theta})$, with parameters estimated, for instance, via a logistic regression. This procedure allows identification of patient profiles significantly associated with the principal stratum, informing the targeted population. 

{Moreover, The PD-Robust framework enables the evaluation of the patient characteristic $Z\in\bfX$ within each of the principal strata defined by $U(t_\ast)$, including the mean $\mu_{Z,U(t_\ast)}\equiv\mathbb{E}[Z|U(t_\ast)]$, the variance $\sigma^2_{Z,U(t_\ast)}\equiv Var[Z|U(t_\ast)]$, and the $\kappa$th quantile $q_{\kappa,U(t_\ast)}\equiv\inf\{q:\mathbb{P}(Z\leq q|U(t_\ast))\geq\kappa\}$. We provide the consistent estimates for the summary statistics of the patient characteristic within the stratum $U(t_\ast)=(1,1)$ in the property below. The detailed derivation, as well as the identification of the summary statistics for the other strata, was summarized in Section $1$ of the Supplementary Material.}

\begin{property}\label{property1}
{Under Assumption $1$-$4$, for any baseline variable $Z\in\bfX$ of the targeted population belonging to the principal stratum $U(t_\ast)$, the mean $\mu_{Z,U(t_\ast)}$, the variance $\sigma^2_{Z,U(t_\ast)}$, and the $\kappa$th quantile $q_{\kappa,U(t_\ast)}$ can be consistently estimated by 
$\hat{\mu}_{Z,U(t_\ast)}=\sum_{i=1}^N\hat{w}_iZ_i$, $\hat{\sigma}^2_{Z,U(t_\ast)}=\sum_{i=1}^N\hat{w}_iZ_i^2-\left[\sum_{i=1}^N\hat{w}_iZ_i\right]^2$, $\hat{q}_{\kappa,U(t_\ast)}=\underset{q}{\arg\min}\hspace{0.1em}(1/N)\sum_{i=1}^N\hat{w}_i\rho_\kappa(Z_i-q)$ for $\kappa\in(0,1)$, respectively, if the PPS model is correctly specified. For the stratum $U(t_\ast)=(1,1)$, we have $\hat{w}_i=p_1(t_\ast,\bfX_i;\hat{\bftheta})/\{\sum_{i=1}^Np_1(t_\ast,\bfX_i;\hat{\bftheta})\}$ 
% Here, we have
% \begin{itemize}
%     \item $\hat{w}_i=p_1(t_\ast,\bfX_i;\hat{\bftheta})/\{\sum_{i=1}^Np_1(t_\ast,\bfX_i;\hat{\bftheta})\}$ for $U(t_\ast)=(1,1)$;
%     \item $\hat{w}_i=[p_0(t_\ast,\bfX_i;\hat{\bftheta})-p_1(t_\ast,\bfX_i;\hat{\bftheta})]/\{\sum_{i=1}^N[p_0(t_\ast,\bfX_i;\hat{\bftheta})-p_1(t_\ast,\bfX_i;\hat{\bftheta})]\}$ for $U(t_\ast)=(0,1)$;
%     \item $\hat{w}_i=1-p_0(t_\ast,\bfX_i;\hat{\bftheta})/\{\sum_{i=1}^N[1-p_0(t_\ast,\bfX_i;\hat{\bftheta})]\}$ for $U(t_\ast)=(0,0)$.
% \end{itemize}
with an estimated parameter $\hat{\bftheta}$. The $\hat{w}_i$ corresponding to the other two strata can be found in the Supplementary Material. The check loss function $\rho_\kappa(\cdot)$ is given by $\rho_\kappa(u)=u(\tau-I\{u<0\})$. }
%   Moreover,  the $\tau$th quantile of $Z$ can be consistently estimated by the estimator $\hat{q}_{\tau,(1,1)}$ defined as
%     \begin{equation*}
% \hat{q}_{\tau,(1,1)}\equiv\underset{q}{\arg\min}\hspace{0.1em}\frac{1}{N}\sum_{i=1}^N\hat{w}_i\rho_\tau(Z_i-q),
%     \end{equation*}
%     where 
\end{property}
By Property \ref{property1}, all summary statistics describing the patient profile of the principal strata can be computed, thereby rendering the studied population tractable in practice.

\section{The impact of HAC on post-fracture outcomes}\label{The case study}

\textbf{Studied cohort and variables.} We evaluated heterogeneous HAC effects on post-fracture recovery outcomes, measured as “days spent at home” (DAH), in older adults living with ADRD following hospital discharge, using Medicare claims data. %Summary statistics for 
The baseline patient-level characteristics before fracture (Figure \ref{introfigure}C) serve as confounders of the impact of HAC on recovery trajectories. More details can be found in the Supplementary Material.

\begin{figure}
    \centering
    \includegraphics[scale=0.27]{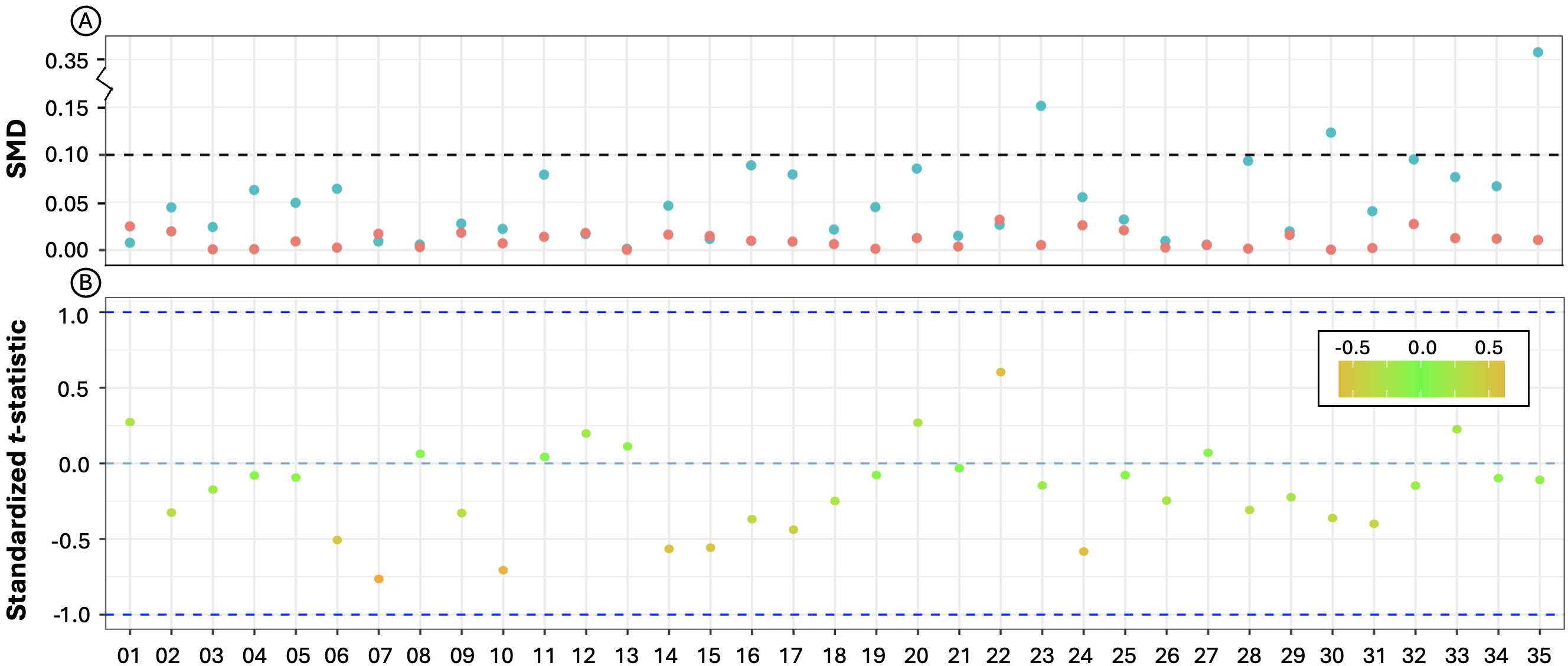}   
     \caption{Evaluation of nuisance models. A: the standardized mean difference (SMD) to evaluate the PS model. Blue points are based on raw data. Red points are based on results after applying inverse probability weighting. B: the standardized t-statistics to evaluate the PPS model. Covariate names are listed in Table S$1$.}
    \label{asptchk}
\end{figure}

\textbf{Diagnostics for nuisance models.} To account for the high mortality rates in our studied cohort (Figure \ref{introfigure}), we applied our PD-Robust method to the principal stratum of patients: those who were alive at month $6$ post-hospital discharge, regardless of exposure. Nuisance functions were fit using GLM for estimating the PS and single time-based CM, while pooled logistic regression was used to estimate the PPS with all confounders (Table S$1$) and/or confounder-by-HAC interactions. {To evaluate the plausibility of the estimated nuisance models. i.e., the PS and the PPS models, we conducted two diagnostic evaluations as described in Section \ref{Practical guidance}. Specifically, we examined whether applying inverse PS weighting reduced the SMD of the baseline covariates. Second, we assessed whether the PPS model achieved the covariate balancing properties outlined in Theorem S$1$ in \cite{jiang2022multiply}. The results are visualized in Figure \ref{asptchk}.  Figure \ref{asptchk}A shows that the SMDs for most covariates were substantially reduced after PS weighting, suggesting that the PS model may be appropriate. Figure \ref{asptchk}B displays the standardized t-statistics for the PCE across all covariates, which should be close to $0$ if the PPS model is correctly specified. As all the t-statistics were within the range of $\pm1$ and centered around $0$, we conclude that the PPS model is likely appropriate.}

\textbf{Potential effect modifiers in HTE working structural models.} Motivated by published findings \citep{orwig2022sex,wang2016electrolyte,shi2024retrospective} suggesting that sex and age could be highly associated with poor recovery outcomes, we considered three sequential HTE working structural models: Model $1$ included only sex (Female/Male); Model $2$ included both sex and age category (Aged under/over $85$ years); and Model $3$ incorporated sex, age category, and the interaction between them.
%sex and age category. 
This sequential modeling strategy allows for a comprehensive assessment of how the HAC effects vary by sex, age, and/or their interaction. 

\begin{figure}
    \centering
        \includegraphics[scale=0.28]{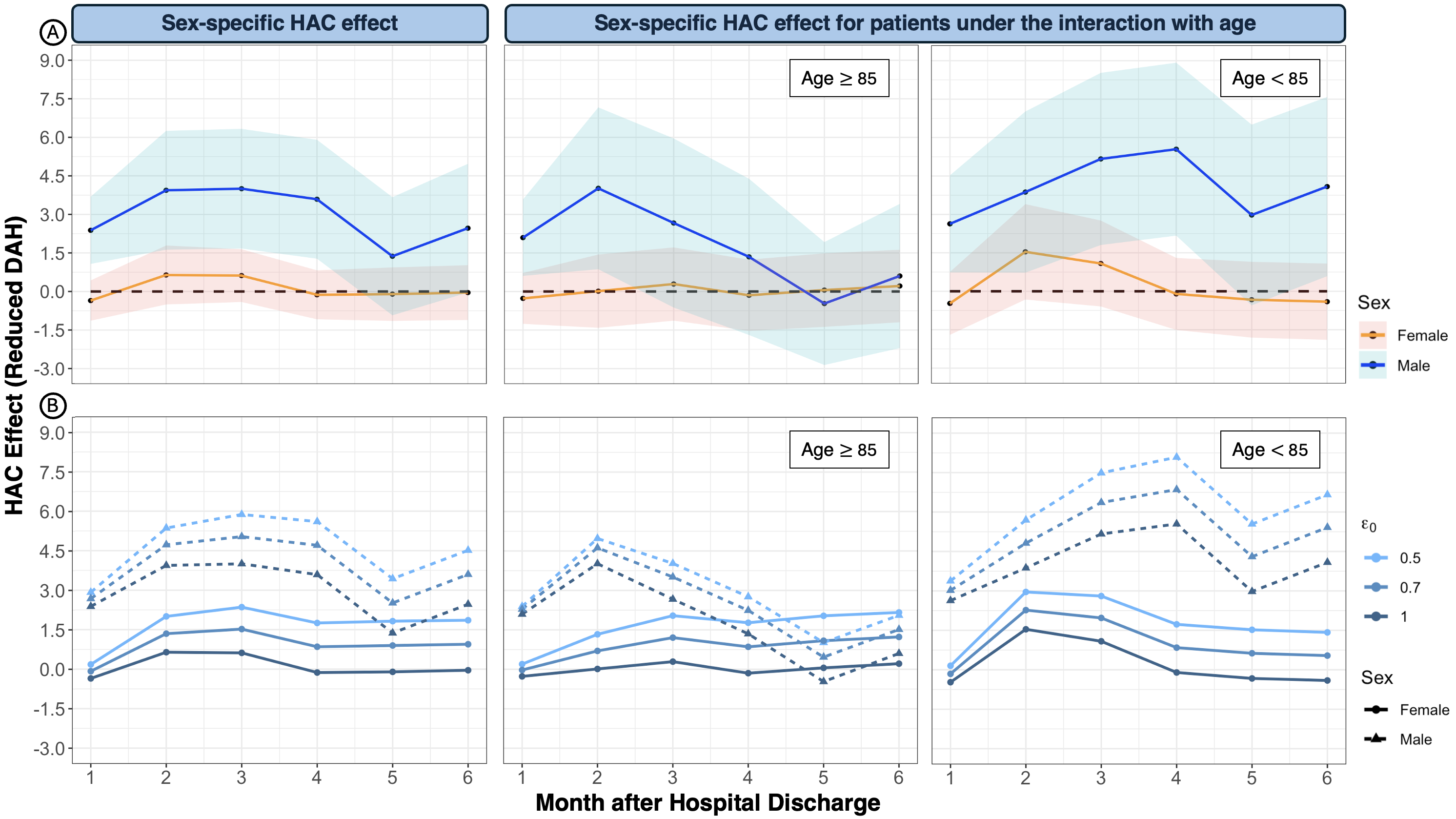}
        \caption{{The estimated HAC effects across six months post discharge. A. Group-specific effects among hip fracture patients who would be alive at month $6$ post discharge regardless of HAC occurrence. Shaded areas represent the $95\%$ confidence region. B. Sensitivity analysis for the violation of principal ignorability assumption under different value of $0<\epsilon_0\leq 1$, with $\epsilon_0=1$ indicating no assumption violation. }} \label{real_alys_monthly}
\end{figure}

\begin{figure}
    \centering
    \includegraphics[scale=0.27]{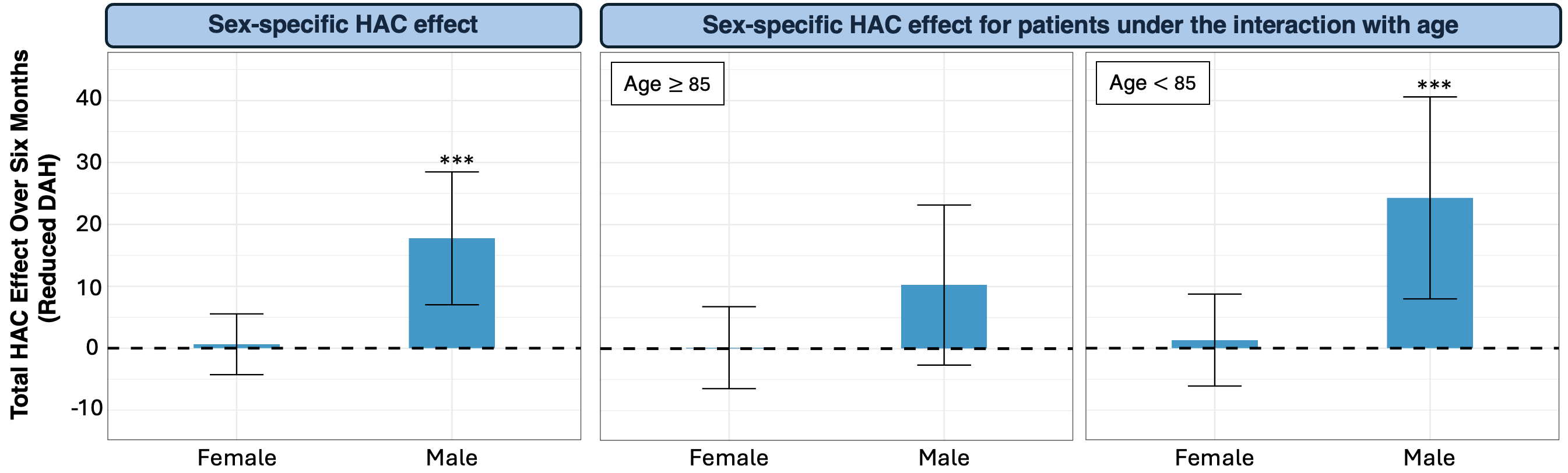}
    \caption{Aggregated HAC effects over six months among patients who would be alive at month $6$ post discharge regardless of HAC occurrence. Error bars represent $95\%$ confidence internals. $\ast\ast\ast: p\leq0.01$ represents statistical significance by comparing PCEs between sex groups.}  \label{real_alys_total}
\end{figure}

\begin{figure}
    \centering
    \includegraphics[scale=0.16]{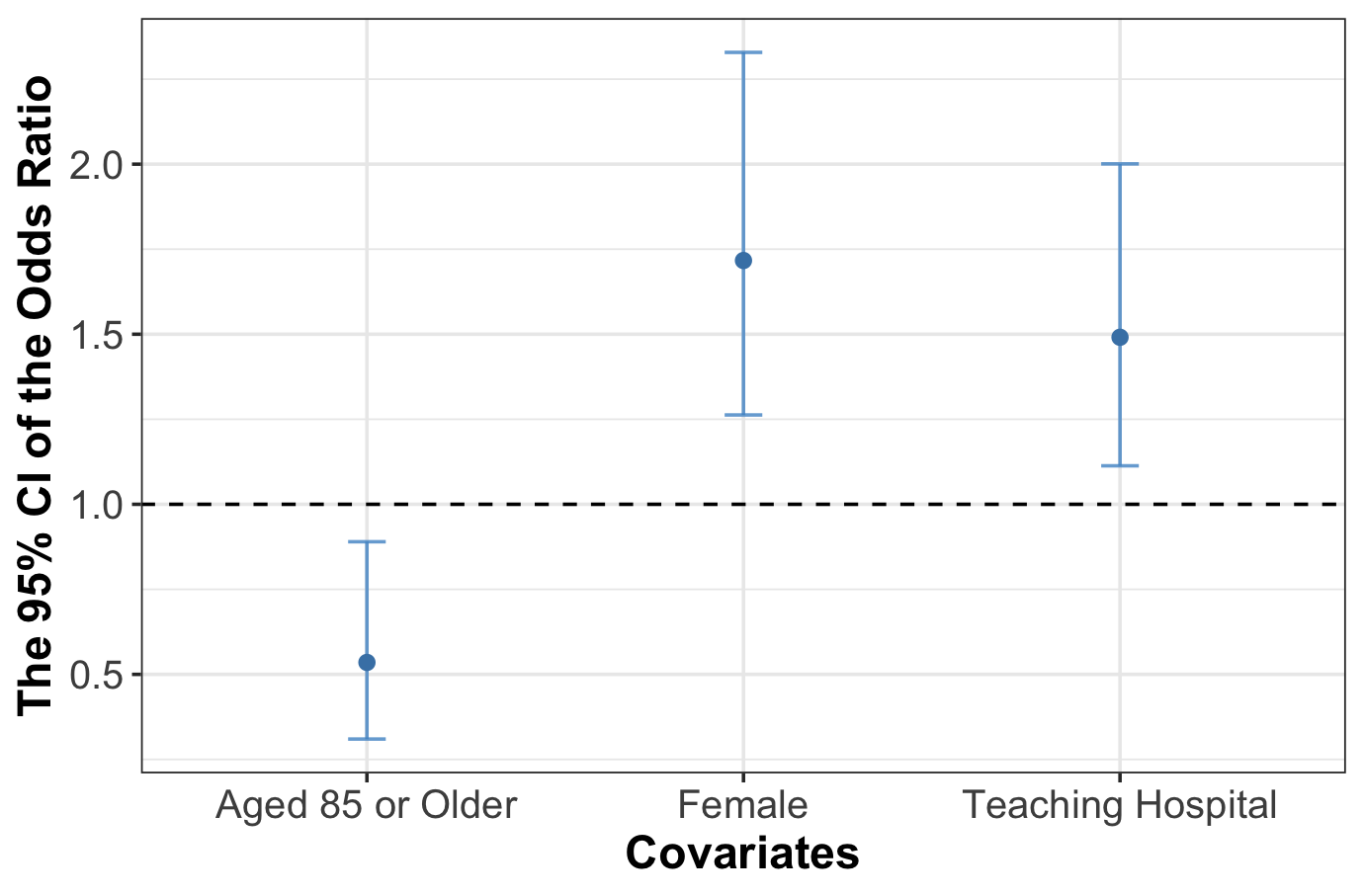}    
    \caption{The $95\%$ confidence interval of the odds ratios for covariates that significantly affect the probability of being alive at month $6$ post discharge regardless of HAC occurrence, after adjusting for other covariates.}
    \label{U11prob}
\end{figure}

\begin{table}[ht!]\footnotesize
\setlength\extrarowheight{-8pt} % default value: 6pt
\centering
\caption{{Comparison of the selected characteristics significantly associated with the principal stratum who would be alive at month $6$ post discharge regardless of the exposure, between the whole population and all three principal strata of patients.}}
      \begin{tabular}{*{6}{c}}
      \toprule
    \midrule
 & &  & \multicolumn{3}{c}{\textbf{Principal Stratum} $U(t_\ast)$}  \\ 
 \cmidrule(l){4-6}\\
 \textbf{Variables: Count (\%)} & &\shortstack{\textbf{Whole Population} \\$(N=21272)$} & \shortstack{$(1,1)$\\$(N\approx14166)$} & \shortstack{$(0,1)$\\$(N\approx2019)$} & \shortstack{$(0,0)$\\$(N\approx5087)$}\\
 \midrule
Age & $65-75$ & $9.6\%$ & $10.8\%$ & $10.8\%$ & $5.7\%$\\ 
 & $76-84$ & $31.0\%$ & $33.1\%$ & $28.0\%$ & $26.3\%$\\ 
 & $\geq85$ & $59.4\%$ & $56.1\%$ & $61.1\%$ & $68.0\%$\\ 
 \midrule
 Sex & Female & $74.9\%$ & $78.2\%$ & $71.5\%$ & $66.8\%$\\
 \midrule
 Teaching Hospital & Yes & $48.2\%$ & $51.7\%$ & $27.4\%$ & $46.8\%$\\
\bottomrule
\end{tabular}%
 \label{u11mean}
\end{table}%

% \begin{table}[ht!]\footnotesize
% \setlength\extrarowheight{-8pt} % default value: 6pt
% \centering
% \caption{Comparison of the selected characteristics significantly associated with the principal stratum between the whole population and the principal stratum of patients who would survive regardless of the exposure at least half a year.}
%       \begin{tabular}{*{6}{c}}
%       \toprule
%     \midrule
%  \textbf{Variables: Count (\%)} & & \textbf{Whole Population} & $U(t_\ast)=(1,1)$ & $U(t_\ast)=(0,1)$&$U(t_\ast)=(0,0)$  \\ 
%  \midrule
% Age & $65-75$ & $9.6\%$ & $10.8\%$ \\ 
%  & $76-84$ & $31.0\%$ & $33.1\%$ \\ 
%  & $\geq85$ & $59.4\%$ & $56.1\%$ \\ 
%  \midrule
%  Sex & Female & $74.9\%$ & $78.2\%$ \\
%  \midrule
%  Teaching Hospital & Yes & $48.2\%$ & $51.7\%$ \\
% \bottomrule
% \end{tabular}%
%  \label{u11mean}
% \end{table}%

\textbf{Sensitivity analyses and post hoc evaluations.} %In addition to 
Along with the primary analysis, we conducted two sensitivity analyses. The first focused on the stratum of patients who survived at least until the tenth months, regardless of exposure. The second assessed the robustness of the resulting estimator to potential {violation of the principal ignorability assumption, as described in Section \ref{Practical guidance}. Besides, we also evaluated the robustness of the estimator to violations of exposure ignorability.} Moreover, we conducted a post hoc evaluation to characterize the potential patient profile from the principal stratum using logistic regression, as detailed in Section \ref{Practical guidance}. 
% This operation is supported by our theoretical result that the PPS equals $\mathbb{P}\left(S(t_\ast)=1|A=0,\bfX\right)$, allowing us to estimate the association between the baseline covariates and membership in the principal stratum, even though the principal stratum itself is not observed.

\textbf{Results.} First, direct observation from HTE models suggests that males may experience larger HAC effects (fewer DAH due to HAC) than females at certain months in all three models (P-value $<0.05$), %Models 1 and 3 (Table S4)
and that the HAC effects among males may also substantially vary in different age categories in Model $3$ (Table S$4$, P-value $<0.05$). To explicitly profile HTE, we visualized the estimated HAC effects across three specific subgroups. Figure \ref{real_alys_monthly}A displays the estimated sex-specific HAC effects with 95\% confidence region over six months after hospital discharge. %with each row representing one principal stratum. 
Key observations include: (\romannumeral 1) HAC appeared to reduce DAH, with males exhibiting larger HAC effects compared to females, whose HAC effects were not significant; (\romannumeral 2) Compared with older patients, HAC effects were greater in patients aged under $85$ years, and became more pronounced for males under $85$ years. 
% This finding suggests that the heterogeneity of HAC effects driven by sex also significantly varied by different age groups. 

Moreover, two sensitivity analyses support the robustness of our findings. First, applying the three models described above to the stratum of patients who were alive at month $10$ revealed slightly higher HAC effects compared to those observed in the current analysis stratum at certain months, although the differences may not be statistically significant (Table S$4$ and Table S$5$). %(Figure S$2$). 
{Second, Figure \ref{real_alys_monthly}B evaluates the robustness of the resulting estimator when the principal ignorability assumption may be violated. The estimated HAC effects remained positive and maintained the overall pattern, indicating that our conclusions are not highly sensitive to the violation of the principal ignorability assumption. Furthermore, Figure S$1$ evaluates how robust the resulting estimator is against the violation of exposure ignorability. The estimated HAC effects slightly varied but maintained the overall pattern, even when the latent variable led to around $20\%$ of the total variance of each nuisance model remaining unexplained. This indicates that our conclusions may not be highly sensitive to moderate levels of exposure ignorability violation. }

The above results consistently imply that experiencing HAC will lead to fewer DAH after hospital discharge on average, and its effects vary across months and patient profiles. The trajectory shapes in Figure \ref{real_alys_monthly} imply a delayed recovery for male patients aged 85 years and older who experienced HAC during hospitalization compared to their counterparts without HAC. In contrast, for male patients younger than 85 years, the adverse effects of HAC appear more sustained, with limited recovery observed even by months $5$ and $6$, indicating a prolonged and potentially cumulative impact of HAC on post-discharge outcomes.  Moreover, Figure \ref{real_alys_total} shows the aggregated HAC effects over six months post-hospital discharge. We observed that, on average, males experienced around $15$ fewer DAH due to HAC compared to females, and males who were younger $85$ years had about $13$ fewer DAH compared to their counterparts who were older than $85$ years.

%For both males and females, HAC effects became significantly greater than zero in individuals aged 75 to 84 who also experienced electrolyte disorders before fracture. Notably, the estimated effect size for this specific subgroup indicated a reduction of $15$ to $25$ days spent healthy at home within six months after hospital discharge, representing a reduction that is both statistically and clinically significant.

Lastly, the post hoc evaluation indicates that younger age ($<85$ years), female, and being treated in a teaching hospital were significantly associated with the principal stratum of interest (Figure \ref{U11prob}, Table \ref{u11mean}). It
%This result 
provides valuable guidance to potentially identify the sub-cohort of patients who belong to the principal stratum for clinical decision making. {To provide useful clinical context, we also added descriptive summaries regarding these key characteristics of the other two strata, i.e., the strata $U(t_\ast)=(0,1)$ and $U(t_\ast)=(0,0)$. We observed in Table \ref{u11mean} that more than $65\%$ patients belonged to the always-survivor stratum, exceeding the proportions in the other two strata. } Besides, {we adapted both the DR-learner and the R-learner to the observed data to estimate the same finite dimensional parameters \citep{morzywolek2023weighted}}. The {adapted} DR-learner yielded results that were similar to those obtained from the PD-Robust method, while the {adapted} R-learner produced more distinct estimates (Table S$6$).

\section{Numerical experiment}\label{Numerical experiment}
\subsection{Data generation}
This section aims to evaluate the performance of the PD-Robust estimation procedure. To this end, we mimicked our case study and generated outcomes with the measurement of a binary exposure and survival status, continuous outcome, and a mixture of both continuous and binary baseline covariates. Specifically, we considered a covariate vector containing ten subject-level confounders $\bfX=(X_1,\ldots,X_{10})^T$, where $X_1,\ldots,X_5$ followed a multivariate normal distribution with mean zeros and variance-covariance matrix $\bfSigma$, and
$X_6,\ldots,X_{10}$ followed a multivariate Bernoulli distribution with means $0.5$ and variance matrix $\bfrho$. Detailed formulas for $\bfSigma$ and $\bfrho$ are displayed in Section S3.2 of the Supplementary Material.  Furthermore, we generated a binary exposure $A$ by a conditional distribution $A|\bfX\sim\text{Bernoulli}\{\pi(\bfX)\}$,  where 
$\text{logit}(\pi(\bfX))= -(1.71 + 0.8X_1 -0.4X_2+ 0.6X_3 -0.5X_4+ 0.7X_5+ 0.1X_6-0.2X_7 +0.3X_8 -0.4X_9+ 0.5X_{10}).$ {Since we observed that almost all patients were alive at month $1$ in our case study, we set the PPS model equal to one at $t=1$ to mimic the real data. However, the proposed method is applicable more broadly and is not limited to this special case.} When $1<t\leq T$, $S^a(t)$ given $S^a(t-1)=1$ was generated by a conditional distribution $S^a(t)|\{\bfX,a,S^a(t-1)=1\}\sim\text{Bernoulli}\{p(\bfX,a)\}$ where $\text{logit}(p(\bfX,a))=b + 2.2X_1 +2.1X_2+1.7X_3 +2.0X_4+ 2.1X_5+1.2X_6+2.2X_7 +2.3X_8 +1.1X_9+ 2.2X_{10}-1.6a.$ We considered low death $(15\%)$ and high death $(30\%)$ cumulative incidence scenarios at the last time point by setting $b=3.6$ and $b=2.6$, respectively. The potential outcomes $Y^{a=1}(t)$ and $Y^{a=0}(t)$ were generated by a conditional Normal distribution $Y^a(t)|\{\bfX,a,t,S^a(t)\}\sim\mathcal{N}(\mu(\bfX,a,t,S^a(t)),3)$, where $\mu(\bfX,a,t,S^a(t))=3.0+(0.4-0.2a)X_1+ (0.5-0.2a)X_2 +(0.1-0.2a)X_3 +(0.9-0.2a)X_4 +(0.4-0.2a)X_5 -(0.2+0.2a)X_6+ (0.8-0.2a)X_7+ (0.7-0.2a)X_8 -(0.2+0.2a)X_9+ (0.5-0.2a)X_{10} + 2S^a(t)-0.5a+  (0.9-0.5a)t$. The observed outcome $Y(t)=AY^{a=1}(t)+(1-A)Y^{a=0}(t)$. {More details of data generating process are provided in Section $3$ of Supplementary Material.} The underlying value $\bfbeta_{t0}$ is obtained by plugging $Y^{a=1}(t)$ and $Y^{a=0}(t)$ in Equation (\ref{def of working fct}) to solve the equation in each Monte Carlo run and averaging all the solutions. In this simulation, we conducted $500$ Monte Carlo runs. For each run, we set $T=6$ and considered sample sizes $N \approx 1500$ and $N \approx 5500$. To demonstrate the triply robust property, we applied generalized linear models (GLM) to estimate CM and PS at each time and pooled logistic regression over time to estimate PPS, considering both correctly-specified or mis-specified models.

% To generate the underlying quantities in (\ref{true ratio}), we generated ?? pairs of counterfactual outcomes using the above generation mechanism and regressed on the space of $\bfX$ to obtain parameters $\bftheta_{t,t_\ast}=(\bftheta_{t,t_\ast,1},\bftheta_{t,t_\ast,2})$. In this simulation, we considered two projection functions specified in (\ref{working functions}) with $\bfeta_1^T(t,t_\ast,\tilde{\bfx})\bftheta_{t,t_\ast,1}=\mathbf X\bftheta_{t,t_\ast,1}$ and $\bfeta_2^T(t_\ast,\tilde{\bfx})\bfbeta_{\ast\ast}=\mathbf X\bftheta_{t_\ast,2}$.

\subsection{Evaluation criteria and result interpretation}

We considered a linear function form $f(\tilde\bfX,t,t_\ast;\bfbeta_t)=\tilde{\bfX}^T\bfbeta_t$ with $\bfg(\tilde\bfX,t,t_\ast)=\tilde\bfX$ for evaluation of PD-Robust, where $\tilde{\bfX}=(1,X_1,X_2,X_3,X_6,X_7)^T\subset\bfX$ is the subset of the covariates, and $\bfbeta_t\in\mathbb{R}^6$ is the parameter vector.
% However, it is still interesting to compare these methods with ours and to assess the extent of deviation from our interested estimand caused by ignoring death and relying solely on observed data.
 The performance of these models was evaluated based on bias and Monte Carlo variance. We evaluated bootstrapped standard deviations and $95\%$ coverage probability for our proposed method. {We also compared our method by adapting the DR-learner \citep{nie2021quasi} and the R-learner \citep{kennedy2023towards} to estimate the same finite-dimensional parameters \citep{morzywolek2023weighted}. It is important to note that the original R-learner and DR-learner focused on infinite-dimensional estimation and inference} and were not  designed to address data with truncation by death, making them theoretically unsuitable for our specific setting. Nevertheless, we included comparisons between the proposed PD-Robust and both the adapted DR-learner and R-learner to examine how finite-dimensional estimation may be influenced by death. For brevity, we only present the results for the first element of $\bfbeta_t$. Other results can be found in Section S$2$ of the Supplementary Material.

\begin{figure}
    \centering
    \includegraphics[scale=0.23]{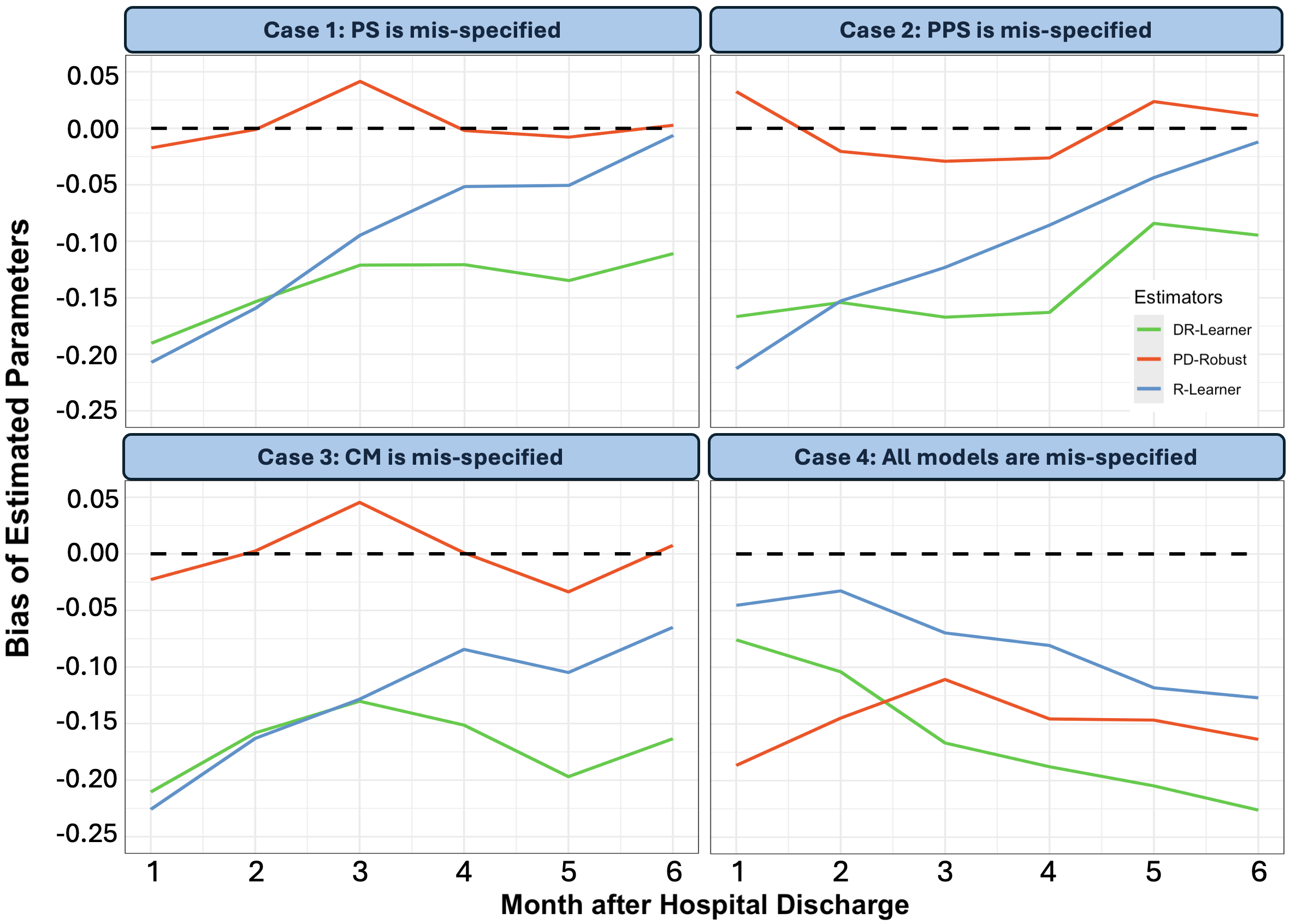}
    % \caption{(?more figures will be added soon?) The evaluation of estimation bias for the first element of $\bfbeta$ over time under four cases due to different levels of model mis-specification ($N\approx 1000$, death rate $\approx 20\%$). The red solid and black dashed lines represent the estimated and true parameters, respectively. The overlapping lines imply unbiased estimation. Shaded areas represent the estimation variability corresponding to one Monte Carlo standard deviation.}
    \caption{The comparison of estimation bias between three estimators for the first element of $\bfbeta_t$ over time under four levels of model mis-specification ($N\approx 5500$, death rate $\approx 30\%$). The red, green, and blue solid lines represent the proposed PD-Robust method, the DR-Learner, and the R-Learner, respectively. The black dashed line represents zero bias.}
    \label{biasplot}
\end{figure}

\begin{table}[ht!]
\setlength\extrarowheight{-8pt} % default value: 6pt
%\centering
\footnotesize
    \centering
    \caption{The evaluation of bootstrap standard deviations (BSD) and the $95\%$ coverage probabilities (CP) of the PD-Robust estimates for the first element of $\bfbeta_t$, under death rate $\approx 30\%$ with $N\approx 5500$. MCSD: Monte Carlo standard devision.}
      \begin{tabular}{@{}  l *{12}{c} @{}}
      \toprule
     &\multicolumn{3}{c}{Case 1 (Wrong PS)} 
     &\multicolumn{3}{c@{}}{Case 2 (Wrong PPS)}
     &\multicolumn{3}{c}{Case 3 (Wrong CM)} 
     &\multicolumn{3}{c@{}}{Case 4 (All wrong)}\\
      \cmidrule(lr){2-4}\cmidrule(l){5-7}\cmidrule(lr){8-10}\cmidrule(l){11-13}
   t &   MCSD & BSD & CP & MCSD & BSD & CP & MCSD & BSD & CP & MCSD & BSD & CP\\
    \midrule
      1  & 0.440 & 0.448 & 0.96 & 0.548 & 0.555 & 0.95 & 0.571 & 0.587 & 0.95 & 0.540 & 0.522 & 0.94\\
      2  & 0.469 & 0.466 & 0.93 & 0.550 & 0.561 & 0.95 & 0.607 & 0.585 & 0.96 & 0.546 & 0.513 & 0.92\\
      3  & 0.493 & 0.464 & 0.94 & 0.590 & 0.558 & 0.93 & 0.609 & 0.585 & 0.94 & 0.529 & 0.513 & 0.92\\
      4  & 0.464 & 0.472 & 0.96 & 0.595 & 0.550 & 0.93 & 0.593 & 0.587 & 0.95 & 0.524 & 0.511 & 0.94\\
      5  & 0.508 & 0.482 & 0.96 & 0.593 & 0.562 & 0.93 & 0.622 & 0.588 & 0.94 & 0.494 & 0.510 & 0.95\\
      6  & 0.472 & 0.489 & 0.95 & 0.583 & 0.554 & 0.94 & 0.573 & 0.581 & 0.95 & 0.498 & 0.505 & 0.94\\
     \bottomrule
      \end{tabular}%
      \label{sdcptb}
\end{table}%

Figure \ref{biasplot} visualizes the bias for four cases under the setting of $N\approx 5500$ and death rate $\approx 30\%$: any two of three nuisance models $\pi(\bfX;\bfalpha)$, $e_{(1,1)}(t_\ast,\bfX;\bftheta)$, and $\mu_{as}(t,t_\ast,\bfX;\bfgamma)$ were correctly specified or all nuisance models were mis-specified due to the missing important covariates, with different death rates and sample sizes. Details of mis-specified models can be found in Section S$3.2$ of the Supplementary Material. We observed that when any two of the three nuisance models were correctly specified, PD-Robust produced unbiased estimates of the underlying truth. In contrast, the two existing approaches that relied solely on the observed data without addressing the issue of death exhibited substantial estimation bias. This highlights the triply robust property of PD-Robust and underscores the importance of properly accounting for death in practice. However, when all models were mis-specified, all estimates showed bias. Moreover, Table \ref{sdcptb} summarizes the Monte Carlo variance and the bootstrap variance of the PD-Robust estimates, along with the coverage probabilities based on the $95\%$ confidence interval. {We observed that two variances were closely aligned, and the coverage probabilities approached the $95\%$ nominal level across all scenarios.} These findings support the use of the bootstrap approach in practice for reliable statistical inference. Note that the CP remained close to its nominal level for time points $4,5,6$ even when all models were mis-specified. This is likely because the estimation variability outweighed the estimation bias. {As shown in our additional simulation study, the coverage probability (CP) decreased with increasing sample size (the estimation variability became smaller). We refer interested readers to Section S$2$ and Table S$12$ of the Supplementary Material for further details.}

{Table \ref{quantiletb} summarizes the results of the estimated means and quantiles (0.25, 0.5, 0,75) of covariates within the principal stratum, under the setting of $N\approx5500$ and death rate $\approx 30\%$. The small magnitude of the absolute bias and the standard deviation indicated that, although the principal stratum is unobserved, our method successfully captures the summary statistics of the covariates within the principal stratum.}

\begin{table}[ht!]
\setlength\extrarowheight{-8pt} % default value: 6pt
%\centering
\small
    \centering
    \caption{The absolute bias and Monte Carlo standard deviation (MCSD) of the estimated means and quantiles (0.25, 0.5, and 0.75) of the covariates within the principal stratum ($30\%$ death rate, $N\approx5500$). The quantiles are not presented for binary variables ($X_6-X_{10}$).}
      \label{quantiletb}
      \begin{tabular}{@{}  l *{8}{c} @{}}
      \toprule
      \midrule
     &\multicolumn{2}{c}{Mean} 
     &\multicolumn{2}{c@{}}{0.25 quantile}
     &\multicolumn{2}{c}{0.5 quantile} 
     &\multicolumn{2}{c@{}}{0.75 quantile}\\
      \cmidrule(lr){2-3}\cmidrule(l){4-5}\cmidrule(lr){6-7}\cmidrule(l){8-9}
   Covariate &  Bias & MCSD & Bias & MCSD & Bias & MCSD & Bias & MCSD \\
    \midrule
      $X_1$  & 0.011 & 0.018 & 0.013 & 0.024 & 0.012 & 0.021 & 0.008 & 0.019\\
      $X_2$  & 0.007 & 0.026 & 0.008 & 0.033 & 0.006 & 0.028 & 0.007 & 0.025\\
      $X_3$  & 0.009 & 0.025 & 0.010 & 0.025 & 0.008 & 0.023 & 0.007 & 0.022\\
      $X_4$  & 0.007 & 0.026 & 0.008 & 0.033 & 0.008 & 0.028 & 0.007 & 0.024\\
      $X_5$  & 0.009 & 0.018 & 0.012 & 0.024 & 0.009 & 0.020 & 0.008 & 0.018\\
      $X_6$  & 0.002 & 0.012 & $-$ & $-$ & $-$ & $-$ & $-$ & $-$ \\
      $X_7$  & 0.002 & 0.012 & $-$ & $-$ & $-$ & $-$ & $-$ & $-$ \\
      $X_8$  & 0.002 & 0.012 & $-$ & $-$ & $-$ & $-$ & $-$ & $-$ \\
      $X_9$  & 0.001 & 0.012 & $-$ & $-$ & $-$ & $-$ & $-$ & $-$ \\
      $X_{10}$  & 0.003 & 0.012 & $-$ & $-$ & $-$ & $-$ & $-$ & $-$ \\
     \bottomrule
      \end{tabular}%
\end{table}%

In addition, similar conclusions hold for other parameter estimates and across alternative data generation settings, including scenarios with $N\approx1500$ and death rate $\approx15\%$ (Tables S$7$-S$14$). {Besides, we also conducted additional simulation where we estimated the nuisance functions by machine learning methods. We observed in Table S$15$ that the proposed estimator yielded small estimation bias under a relatively large sample size.} Collectively, these results support the validity of the PD-Robust method.

\section{Discussion and conclusion}\label{Discussion}
We have developed a new PD-Robust framework to robustly assess effect modification on an outcome trajectory in the presence of truncation by death under principal ignorability. {The proposed new estimand is build upon projection techniques and an interpretable structural working model focusing on finite dimensional parameter estimation. This framework also facilitates model diagnosis, several sensitivity analysis strategies for evaluating the impact
of assumption violations, and} the characterization of patient profiles among the principal stratum, a desirable feature in practice.  Both theoretical analysis and simulation studies have demonstrated the validity of the PD-Robust method, showcasing its robustness against estimation bias, accurate statistical inference, and efficient computational performance.

Applying this method, we identified significant HAC effects on post-fracture DAH after hospital discharge, with effect sizes varying across patient profiles. Notably, recent literature has highlighted that males experience higher mortality after hip fractures and exhibit significantly different functional outcomes compared to females in terms of disability, gait speed, and depressive symptoms \citep{orwig2022sex,mutchie2023associations,mehta2024association}. Our analysis further contributes to understanding sex difference in post-fracture DAH trajectories—a patient-centered outcome that reflects the extent to which individuals live safely, independently, and comfortably in their homes and communities. We identified sex as a potential effect modifier of HAC, with pronounced and significantly reduced DAH (approximately $23$ fewer days within six months post-discharge) in males under $85$ years comparing HAC to no HAC during hospitalization. This reduction is both statistically and clinically significant, especially in light of recent clinical literature suggesting that a difference greater than eight DAH is considered clinically meaningful \citep{auriemma2023stakeholder}. In contrast, female patients appeared more robust to HAC exposure, with estimated effects near zero across age groups. These findings add to the current understanding of recovery differences and highlight high-risk sub-cohorts with HAC stratified by sex, age, and their interaction, emphasizing the need for health systems to prioritize these patients with tailored strategies to improve recovery outcomes and support aging in place. 

Moreover, the comparison among different methods revealed similar results between the proposed PD-Robust and the adapted DR-learner, the latter of which relies solely on observed data. This finding may appear counterintuitive, but it could be attributed to the fact that, in this case study, the principal stratum does not substantially deviate from the observed population remaining alive (Table \ref{u11mean}). In addition, the pre-defined value of $t_\ast$ for the principal stratum may influence the results. As indicated in Section \ref{The case study}, using $t_\ast=10$ yielded slightly different HAC effects compared to $t_\ast=6$, resulting in greater deviation from the adapted DR-learner estimators at certain months. Nevertheless, we recommend the use of PD-Robust in broader applications, as it more reasonably accounts for death as an important consequence without assuming censoring at random, an especially critical consideration when death rates are high or when the exposure is believed to differentially influence mortality across patient profiles. In practice, larger $t_\ast$ defines a more distinct principal stratum but possibly less directly related to the impact of baseline exposures and covariates of interest. We recommend end-users rely on relevant clinical knowledge to guide the selection of $t_\ast$. Further discussion of method extensions can be found in Section S$2$ of the Supplementary Material.

Despite its advancements, the PD-Robust estimation method has several limitations that warrant future investigation: (\romannumeral 1) the current approach focuses on a difference-based estimand, rather than other effect measures such as relative risk or odds ratios; (\romannumeral 2) the method could be sensitive to extreme values of inverse probability weighting, potentially leading to instability in estimation—trimming may help mitigate this issue; (\romannumeral 3) the positivity assumption may not always hold in real-world scenarios, and extending the method to focus on treated populations or overlapping populations under clinical equipoise is an interesting area; {(\romannumeral 4) although Monotonicity assumption is plausible in our application, it may not hold in more general settings. Sensitivity analyses are therefore needed to assess the impact of potential violations of this assumption. Moreover, the exposure ignorability and principal ignorability assumptions could be relaxed through partial identification or bounding approaches. These would be valuable directions for future research, although beyond the
scope of the current paper;} and (\romannumeral 5) the method currently addresses binary exposures, which aligns with our real data application. However, generalizing PD-Robust to handle continuous exposures (e.g., duration of surgery) presents a methodologically interesting challenge. 

\bibliographystyle{imsart-nameyear} % Style BST file
\bibliography{reference}       % Bibliography file (usually '*.bib')

%% or include bibliography directly:
% \begin{thebibliography}{}
% \bibitem[\protect\citeauthoryear{???}{???}]{b1}
% \end{thebibliography}

\end{document}